\documentclass[a4paper]{article}
\pdfoutput=1 
\PassOptionsToPackage{%
backend=biber, 
language=auto,%
style=numeric-comp,%
sorting=none, 
maxbibnames=3, 
natbib=true 
}{biblatex}
\usepackage{biblatex}

\usepackage{fontawesome}
\newcommand{\github}[1]{%
   \href{#1}{\faGithubSquare}%
}

\usepackage{silence}
\usepackage{jcappub} 
\usepackage{float}

\usepackage[T1]{fontenc} 

\usepackage{dcolumn}
    \newcolumntype{d}[1]{D{.}{\cdot}{#1}}
    \newcolumntype{.}{D{.}{.}{-2}}
    \newcolumntype{,}{D{,}{,}{-3}}

\usepackage{booktabs}

\usepackage{soul}
\usepackage{hhline}
\usepackage{arydshln}
\usepackage{physics}

\usepackage{booktabs}

\def\NHUNIT{\ifmmode {\rm \,cm^{-2}} \else $\rm \,cm^{-2}$ \fi} 
\def\clee{$C_\ell^{EE}$}
\def\clbb{$C_\ell^{BB}$}

\def\muKcmb{\ifmmode \,\mu$K$_{\rm CMB}$\else \,$\mu$K$_{\rm CMB}$\fi}

\newcommand{\ch}{\mathrm{ch}}

\newcommand{\nchs}{n_\ch}
\newcommand{\na}{\mathrm{A}}
\newcommand{\nb}{\mathrm{B}}
\newcommand{\nc}{\mathrm{C}}

\newcommand{\ncross}{\mathrm{cross}}

\newcommand{\axa}{{\na\times\na}}
\newcommand{\axb}{{\na\times\nb}}
\newcommand{\axc}{{\na\times\nc}}
\newcommand{\bxb}{{\nb\times\nb}}
\newcommand{\bxc}{{\nb\times\nc}}
\newcommand{\cxc}{{\nc\times\nc}}

\newcommand{\As}{A_{\rm s}}

\newcommand{\ns}{n_{\rm s}}
\newcommand{\lcdm}{$\Lambda$CDM}
\newcommand{\rpivot}{r}
\newcommand{\lcdmr}{$\Lambda$CDM + $\rpivot$}

\newcommand{\omegab}{\Omega_{\rm b}}
\newcommand{\omegac}{\Omega_{\rm cdm}}
\newcommand{\hnot}{H_{\rm 0}}

\newcommand{\OmegaM}{\ifmmode\Omega_{\rm M}\else $\Omega_{\rm M}$\fi}

\newcommand{\mnu}{\sum m_\nu}

\providecommand{\Planck}{\textit{Planck}}

\providecommand{\text}[1]{\rm{#1}}

\newcommand{\begm}{\begin{pmatrix}}
\newcommand{\enm}{\end{pmatrix}}

\providecommand{\Tr}{\text{Tr}}

\def\pmb#1{\setbox0=\hbox{#1}%
    \kern-.025em\copy0\kern-\wd0
    \kern.05em\copy0\kern-\wd0
    \kern-.025em\raise.0433em\box0}

\def\p2Y{\;_2Y}
\def\m2Y{\;_{-2}Y}

\newcommand{\mksym}[1]{\ifmmode {\rm #1}\else #1\fi}

\providecommand{\text}[1]{\rm{#1}}

\providecommand{\CAMB}{\texttt{CAMB}}

\newcommand\ba{\begin{eqnarray}}
\newcommand\ea{\end{eqnarray}}
\newcommand\bea{\begin{eqnarray}}
\newcommand\eea{\end{eqnarray}}

\newcommand\be{\begin{equation}}
\newcommand\ee{\end{equation}}

\newcommand{\fsky}{\ensuremath{f_{\rm sky}}}

\providecommand{\Tr}{\text{Tr}}

\title{Accurate and efficient likelihood modeling for large-scale CMB data}

\author[a,b,1]{Giacomo Galloni,\note{Corresponding author.}}
\author[b,c]{Paolo Campeti,}
\author[a,b,d]{Luca Pagano,}
\author[b]{Martina Gerbino,}
\author[b]{Massimiliano Lattanzi}
\author[a,b]{and Paolo Natoli}

\affiliation[a]{Dipartimento di Fisica e Scienze della Terra, Università degli Studi di Ferrara, via Saragat 1, I-44122 Ferrara, Italy}
\affiliation[b]{Istituto Nazionale di Fisica Nucleare, Sezione di Ferrara, via Saragat 1, I-44122 Ferrara, Italy}
\affiliation[c]{ICSC, Centro Nazionale “High Performance Computing, Big Data and Quantum Computing”}
\affiliation[d]{Institut d'Astrophysique Spatiale, CNRS, Univ. Paris-Sud, Universit\'{e} Paris-Saclay, B\^{a}t. 121, 91405 Orsay cedex, France}

\emailAdd{giacomo.galloni@unife.it}

\abstract{
Accurate parameter estimation from cosmic microwave background data requires reliable likelihood modeling, particularly at large angular scales where angular power spectrum estimators exhibit non-Gaussian statistics. We present a novel approach, based on the Hamimeche-Lewis formalism, that marginalizes over auto-spectra, thus reducing residual biases from noise misestimation and partial sky coverage. We validate our approach by simulating three independent CMB channels, or data splits, in a multi-field setting, comparing to the pixel-based likelihood ground truth estimates for the optical depth $\tau$ and the tensor-to-scalar ratio $r$. We benchmark our method against the main power spectrum-based alternatives available in the literature, showing that it outperforms all of them in terms of accuracy, while remaining fast and computationally efficient \href{https://github.com/ggalloni/CHARM-Like}{\color{blue}\faGithub}.}

\arxivnumber{2505.24829}

\begin{document}
\ActivateWarningFilters[warn]

\maketitle

\flushbottom

\section{Introduction}\label{sec:introduction}

The Cosmic Microwave Background (CMB) is a cornerstone of modern cosmology, offering deep insights into the early Universe. Its anisotropies are thought to originate from quantum fluctuations during inflation, which seeded small inhomogeneities in the primordial plasma \cite{seljakSignatureGravityWaves1997, kamionkowskiStatisticsCosmicMicrowave1997, kamionkowskiQuestModesInflationary2016}.

As CMB photons travel through the Universe, they acquire additional imprints from various processes \cite{Hu:1995em, EISENSTEIN2005360, Eisenstein_2005, Sachs_1967, Rees, Vishniac_1987}. In particular, polarization—typically decomposed into curl-free ``E-mode'' and divergence-free ``B-mode'' components—serves as a key observable. While scalar perturbations only generate E-modes, inflation-driven tensor perturbations would also produce a primordial B-mode signal \cite{Guzzetti:2016mkm, seljakLineofSightIntegrationApproach1996, huWanderingBackgroundCMB1995} that remains unobserved \cite{tristram2022ImprovedLimitsTensortoscalarratio, galloniUpdatedConstraintsAmplitude2022, galloni2024RobustConstraintsTensorPerturbations}.

A wide range of experiments has been deployed to study the CMB, or will be in the near future. Full-sky observations, unique to satellites \cite{boggessCOBEMissionIts1992, wmap_2013, Planck_parameters, Planck_like, Hazumi:2019lys, PTEP_LiteBIRD}, are especially important when studying large-scale signals, such as the potential imprint of primordial gravitational waves (GWs), or the reionization epoch \cite{Ade_2014, collaboration2021BICEPKeckXIIIImproved, galloniUpdatedConstraintsAmplitude2022, Pagano_2020, Natale_2020, Lattanzi_2017, de_Belsunce_2021}.

In a full-sky scenario, assuming that the fields (pure CMB and noise) observed are Gaussian and statistically isotropic, the likelihood of the observed data given a theoretical model is exactly described by a Wishart distribution of the angular power spectra, the $C_\ell$ \cite{hamimeche2008LikelihoodAnalysisCMBTemperature, gerbinoLikelihoodMethodsCMB2020}. However, even satellites which can access the full sky must contend with bright foreground signals that prevent direct complete CMB imaging (see e.g. \cite{PTEP_LiteBIRD, Remazeilles2024MappingHotGasLiteBIRD, Fuskeland2023ExtentedFrequencyConfigurationLiteBIRD}). Multi-frequency observations allow for foreground separation, but this process is not perfect \cite{stompor2009MaximumLikelihoodAlgorithm, stolyarov2002AllskyComponentSeparation, eriksen2008JointBayesianComponent, delabrouille2003MultidetectorMulticomponentSpectral, delabrouille2009FullSkyLow, bennett2003FirstYearWilkinsonMicrowave, caronesAnalysisNILCPerformance2022, carones2023MultiClusteringNeedletILCCMB}. Thus, it is reasonable to assume a partial sky coverage—a cut-sky—for future CMB experiments.

This creates a statistical challenge in how to model the probability of the data at large angular scales. In principle, one could work at the map level, assuming that pixel values follow a multivariate Gaussian distribution \cite{gerbinoLikelihoodMethodsCMB2020}. Various examples of this approach can be found in the literature, e.g. the analysis of COBE satellite \cite{smoot1992cosmicBackgroundExplorer, smootCOBEObservationsResults1999}, where multiple frequency channels were combined at map-level, filtering non-stochastic real-space components \cite{Gorski:1994uu, Gorski:1994ye, Gorski:1996cf, Gorski:1996ti}.

However, this approach quickly becomes computationally expensive and is highly sensitive to systematics, as it depends on the ability to estimate potentially large pixel covariance matrices \cite{hamimeche2008LikelihoodAnalysisCMBTemperature}. Other approaches to overcome this issue are based on sampling the full joint CMB posterior, e.g. through Gibbs sampling \cite{Jewell:2008hg, Taylor:2007sj, Eriksen:2004ss}.
Instead, since the angular power spectrum provides a sufficient summary statistic of cosmological information in the map, harmonic-space likelihoods are commonly used \cite{Percival_2006, Tegmark_2001}.

One such approach in the cut-sky is the Hamimeche \& Lewis (HL) likelihood, which accounts for the non-Gaussianity of $C_\ell$ in full-sky and for couplings between multipoles introduced by masking \cite{hamimeche2008LikelihoodAnalysisCMBTemperature}. Another interesting strength of this approximation is its native capability of working in a ``multi-field'' scenario. In fact, it can be applied in joint analysis across correlated fields, which could be CMB temperature and polarization, or equivalently maps observed at different frequencies, or data splits from a mission dataset \cite{hamimeche2008LikelihoodAnalysisCMBTemperature}. Despite being a good approximation in the cut-sky regime and widely used in the literature, the HL likelihood does not fully capture the distortion of the $C_\ell$ distribution, which deviates from a Wishart form when masking is applied. In addition, semi-analytical cross-spectrum-based likelihood approximation schemes were explored in \cite{de_Belsunce_2021, deBelsunce:2022yll} providing interesting alternatives.

In this work, we revisit the HL likelihood in the context of multi-field CMB polarization analyses, focusing on its practical application in realistic settings. In particular, we address two common challenges: partial sky coverage and imperfect knowledge of the noise bias \cite{Delouis2019Sroll2, planck2016reductionLargeScaleSystematics}, both of which affect the performance of this approximation. Indeed, imperfect knowledge of the systematics of an experiment, or of the residual foreground signal mentioned above, could lead to a mismatch between the assumed noise bias and the true one. To this end, we study two variants of the HL prescription. The first is what we call cross-spectra HL (cHL), which tries to extend the offset HL (oHL) already adopted in the literature \cite{mangilli2015LargescaleCMBTemperaturepolarization, tristramm.2021PlanckConstraintsTensortoscalarratio, tristram2022ImprovedLimitsTensortoscalarratio} to the multi-field case \cite{Vanneste:2019jnz, bicepPlanck2015jointAnalysis}. The cHL is designed to use only cross-spectra, i.e. cross-correlations between two different fields, to by-pass the misestimation of the noise bias. The second, which we introduce here for the first time, is a new \textit{marginalized HL} (mHL) approach, in which auto-spectra are marginalized over analytically to effectively remove dependence on uncertain noise levels.

We carry out a systematic comparison of the HL, cHL, and our proposed mHL using synthetic maps, both in full-sky and cut-sky regimes. As noted above, we focus on large angular scales with $\ell \leq 32$, targeting the reionization peak of the E- and B-mode spectra. Although the likelihood on smaller angular scales is not perfectly Gaussian—hence the original motivation for the HL approach \cite{hamimeche2008LikelihoodAnalysisCMBTemperature}, which is validated on such scales—the central limit theorem ensures that the distribution becomes increasingly Gaussian as higher multipoles are considered. In this regime, residual foregrounds or systematic effects typically represent a more significant limitation \cite{PTEP_LiteBIRD, Remazeilles2024MappingHotGasLiteBIRD, Fuskeland2023ExtentedFrequencyConfigurationLiteBIRD}. To explore the limits of the application of each method, we simulate mismatches in the noise bias and assess their impact on parameter inference. To our knowledge, this is the first comprehensive study that jointly investigates the behavior of HL-based likelihoods under simultaneous noise mismatches and sky cuts, and the first to introduce and validate the mHL as a robust alternative. These results provide new insights into the practical use of HL approximations and offer concrete recommendations for future CMB polarization analyses. For example, we study whether a single-field approach is preferable to multi-field HL-based likelihoods.


\section{Modeling large-scale CMB data}

\subsection{The Hamimeche \& Lewis likelihood}\label{sec: hl_like}

On the full sky, the exact likelihood of the angular power spectra of two correlated CMB fields, for example temperature T and E-modes, is \cite{gerbinoLikelihoodMethodsCMB2020}
\begin{equation}
    -2\ln\mathcal{L}(\hat{\mathbf{C}}_\ell|\mathbf{C}_\ell) = \sum_\ell \qty(2\ell+1)\qty(\Tr{\hat{\mathbf{C}}_\ell \mathbf{C}_\ell^{-1}} - \ln\qty|\hat{\mathbf{C}}_\ell \mathbf{C}_\ell^{-1}| - n)\ ,
\label{eq:exact_like}
\end{equation}
where
\begin{equation}
    \mathbf{C}_\ell = \qty(\begin{matrix}
        C_\ell^{TT} & C_\ell^{TE} \\
        C_\ell^{TE} & C_\ell^{EE}
    \end{matrix}),
\label{eq: full_hl_mat}
\end{equation}
$C_\ell$ indicates the true power spectra, the hat indicates their estimates, and $n=2$ is the number of fields used to normalize the likelihood.

From this starting point, we can define the HL likelihood. \autoref{eq:exact_like} can be rewritten as \cite{hamimeche2008LikelihoodAnalysisCMBTemperature}
\begin{equation}
    -2\ln\mathcal{L}(\hat{\mathbf{C}}_\ell|\mathbf{C}_\ell) = \sum_{\ell\ell^\prime} X_{\ell}^T \eval{M^{-1}_{\ell \ell^\prime}}_{\rm fid.} X_{\ell}\ ,
\label{eq:hl_likelihood}
\end{equation}
where $M^{-1}_{\ell \ell^\prime}$ is the inverse covariance matrix of the $n(n+1)/2$ independent $C_\ell$ components, evaluated for a fiducial model. Then, 
\begin{equation}
    X_{\ell} = \qty(X_\ell^{\rm T\times T},~ X_\ell^{\rm T\times E},~ X_\ell^{\rm E\times E})
\end{equation}
is the vector formed from the independent elements of the matrix
$\mathbf{C}_{g\ell} = \mathbf{C}_{f\ell}^{1/2} \mathbf{U}_\ell \mathbf{g}(\mathbf{D}_\ell) \mathbf{U}^T_\ell \mathbf{C}_{f\ell}^{1/2}$. 
Here, $\mathbf{C}_{f\ell}$ is the fiducial version of the matrix in \autoref{eq: full_hl_mat}, 
$\mathbf{U}_\ell$ is the diagonalization matrix of $\mathbf{C}_\ell^{-1/2}\hat{\mathbf{C}}_\ell \mathbf{C}_\ell^{-1/2}$, and 
$\mathbf{D}_\ell$ is the diagonal matrix of eigenvalues. The function acting on it is defined as \cite{hamimeche2008LikelihoodAnalysisCMBTemperature}
\begin{equation}
    \mathbf{g}(\mathbf{D}_\ell) = \qty[\mathbf{g}(\mathbf{D}_\ell)]_{ij} = g(D_{\ell, ii})\delta_{ij} = \text{sign}\qty(D_{\ell, ii} - 1)\sqrt{2\qty(D_{\ell, ii} - \ln D_{\ell, ii} - 1)}\ \delta_{ij}\ .
    \label{eq: g_func}
\end{equation}

Up to now, introducing the HL likelihood, we used the familiar example of CMB temperature and E-mode polarization \cite{gerbinoLikelihoodMethodsCMB2020}. However, this formalism is defined in general for $n$ Gaussian and statistically isotropic correlated fields \cite{hamimeche2008LikelihoodAnalysisCMBTemperature}. For example, we could consider three generic fields $\qty{A, B, C}$, which in principle could be different frequency channels, data-splits, or actual CMB observables such as temperature, E-modes, and B-modes. The only difference between these cases is the degree to which these fields are correlated.

In practice, CMB analyses are often done in such a multi-field framework, either because one wants to include multiple observables, or because one would benefit from the fact that noise is often uncorrelated between splits. An example is the likelihood constructed for the BICEP/Keck analysis \cite{collaboration2021BICEPKeckXIIIImproved}, where eleven maps are combined in a single HL likelihood.

In this work, we consider only B-(E-)modes from $\nchs$ ``channels'' to estimate the tensor-to-scalar ratio $\rpivot$ (optical depth $\tau$), corresponding to $\nchs$ frequency bands observing the same CMB. Each channel is affected by isotropic white noise $N_\ell^{\rm ch}$, which does not correlate between channels. For $\nchs = 3$ channels $\qty{A, B, C}$, the matrix of \autoref{eq: full_hl_mat} reads

\begin{equation}
    \mathbf{C}_\ell = \qty(\begin{matrix}
        C_\ell^\axa + N_\ell^\na & C_\ell^\axb & C_\ell^\axc \\
        C_\ell^\axb & C_\ell^\bxb  + N_\ell^\nb  & C_\ell^\bxc \\
        C_\ell^\axc & C_\ell^\bxc & C_\ell^\cxc + N_\ell^\nc 
    \end{matrix})\ ,
\label{eq:3chs_full_hl_mat}
\end{equation}
and the data vector reads
\begin{equation}
    X_{\ell} = \qty(X_\ell^\axa,~ X_\ell^\axb,~ X_\ell^\axc,~ X_\ell^\bxb,~ X_\ell^\bxc,~ X_\ell^\cxc)
\label{eq: hl_Xell}
\end{equation}

The HL likelihood requires an estimate of the noise bias and assumes a fiducial cosmological model to compute both $X_\ell$ and $M^{-1}_{\ell \ell^\prime}$. In practice, both of these can be misestimated, either because the noise is not perfectly known, or because the fiducial spectrum differs from the true one. Although the second issue can be addressed by iterating the analysis over multiple fiducial models, the noise bias is trickier to correct since the true level may not be directly accessible. For this reason, we show the results of the wrong fiducial assumption in \autoref{app: wrongfid}, as this would not be an issue in a real-life scenario. There we show that the HL is extremely stable against this effect, as expected \cite{hamimeche2008LikelihoodAnalysisCMBTemperature}. mHL yields similar robustness, while the cHL appears more sensitive.

To handle this, a variation called the offset HL (oHL) likelihood was proposed \cite{mangilli2015LargescaleCMBTemperaturepolarization} (\autoref{sec: oHL}). Assuming uncorrelated noise between fields, one can build a likelihood based only on cross-spectra, avoiding the need to subtract noise bias explicitly. The name ``offset'' will be explained in the next section, where we will try to extend this method to multiple fields.

In this work, we propose a new alternative: the \textit{marginalized HL} (mHL) likelihood, where we marginalize the HL likelihood over the auto-spectra to mitigate this problem (\autoref{sec: mHL}).

We perform an extensive validation of these multi-field HL-based likelihoods, providing concrete recommendations for CMB data analysis. We assess their robustness when the noise levels and fiducial spectra used in the likelihood analysis are mismatched. This is done both in the full-sky, where the HL likelihood is exact, and in the cut-sky, where all of them are approximations. In addition, the usage of a multi-field approach can answer a simple, but very important dilemma. As mentioned before, the $\nchs$ maps used here can be seen as data-splits from a ``full-mission'' dataset. Thus, with this full-mission map, we can ask ourselves what would be the corresponding single-field result. What we call from now the ``single-field'' HL is very useful to understand whether a future CMB analysis should rely on a single-field approach, or prefer a multi-field HL-based likelihood at large angular scales.

\subsubsection{Offset Hamimeche \& Lewis} \label{sec: oHL}

Ref.~\cite{mangilli2015LargescaleCMBTemperaturepolarization} proposed a method to handle misestimates of the noise bias using only cross-spectra. In particular, the oHL likelihood was introduced and validated using pairs of frequency channels (e.g., 100 GHz and 143 GHz from the \textit{Planck} satellite \cite{Planck_parameters, Planck_like}), resulting in a single cross-spectrum. For example, $C_\ell^{100\text{x}143}$ is transformed—following the same steps described earlier—into $X_\ell^{100\text{x}143}$, which is then used in the likelihood computation. However, it is evident from the logarithm and square-root of \autoref{eq: g_func} that cross-spectra can be problematic, as they are not necessarily positive-definite. To address this, Ref.~\cite{mangilli2015LargescaleCMBTemperaturepolarization} introduced an offset $O_\ell$, designed to mimic the presence of a noise bias and enable the cross-spectrum to behave like an auto-spectrum (see the appendix of \cite{mangilli2015LargescaleCMBTemperaturepolarization} for further details).

The proposed offset is defined such that more than 99\% of their Monte Carlo (MC) simulations yield a positive-definite matrix $\mathbf{C}_\ell^{-1/2}\hat{\mathbf{C}}_\ell \mathbf{C}_\ell^{-1/2}$. Thus, it depends on the noise levels in the simulated maps, the fiducial spectra $\mathbf{C}_{f\ell}$, and the mask used \cite{mangilli2015LargescaleCMBTemperaturepolarization}.

This offset, however, could be implemented in different ways, for example based on the expected width of the distribution of $C_\ell$ \cite{tristram2022ImprovedLimitsTensortoscalarratio}, bringing about different outcomes. Although this was discussed in Ref.~\cite{mangilli2008ImpactCosmicNeutrinosGravitationalwave} and we give some extra detail in \autoref{app: offset}, we leave a detailed investigation of this dependence to future work. In the following sections, we always adopt the procedure used in \cite{mangilli2015LargescaleCMBTemperaturepolarization}.

To further avoid numerical issues and regularize the transformation in \autoref{eq: g_func}, Ref.~\cite{mangilli2015LargescaleCMBTemperaturepolarization} proposed modifying the function $g$ as
\begin{equation}
    g(x) \longrightarrow \text{sign}(x)g(\abs{x})\ .
\label{eq:g_func_oHL}
\end{equation}

Although this modification has only a minor impact once the offset is applied, it is important to stress that it is introduced solely for numerical stability—not for statistical reasons. As such, if the offset is not large enough to compensate for negative elements, this adjustment could bias the resulting distribution.

\subsubsection{Extending oHL to multiple fields}\label{sec: cHL}

One crucial aspect is still missing: the oHL is proposed for a single cross-spectrum, so it provides no way of combining the information from multiple cross-spectra. In our case, however, we would like to combine three independent cross-spectra: $C_\ell^\axb$, $C_\ell^\axc$, and $C_\ell^\bxc$. Following the logic of Ref.~\cite{mangilli2015LargescaleCMBTemperaturepolarization}, we try to extrapolate a multi-field generalization, which we will name cross-spectra HL (cHL). To define the cHL, one is easily tempted to build the data vector by concatenating these three spectra, ensuring that no auto-spectra are included. Thus, we can define the new data vector as
\begin{equation}
    C_\ell^{\rm cHL} = \qty(C_\ell^\axb + O_\ell^\axb, C_\ell^\axc + O_\ell^\axc, C_\ell^\bxc + O_\ell^\bxc)\ ,
\label{eq: datavec_ohl}
\end{equation}
which is three times as long as $C_\ell^\axb$, for example, and is then transformed using the HL prescription into $X_\ell^{\rm cHL}$ \footnote{Since each cross-spectrum is treated independently when transforming $C_\ell^{\rm cHL}$, this approach effectively corresponds to using matrices like
\begin{equation}
    \mathbf{C}_\ell^{\rm cHL} = \qty(\begin{matrix}
        C_\ell^\axb + O_\ell^\axb & 0 & 0 \\
        0 & C_\ell^\axc + O_\ell^\axc & 0 \\
        0 & 0 & C_\ell^\bxc + O_\ell^\bxc
    \end{matrix})
\label{eq: ohl_mat}
\end{equation}
in the HL procedure (compare this with \autoref{eq:3chs_full_hl_mat}).}. Then, the procedure is the same as in \autoref{sec: hl_like}. Consider as an example the $X_\ell^\axb$ element of the HL from \autoref{eq: hl_Xell}. Given that the data vector $X_\ell^{\rm cHL}$ is obtained through a single-field HL transformation of \autoref{eq: datavec_ohl}, we note that $X_\ell^{\axb, \rm cHL} \neq X_\ell^{\axb, \rm full~HL}$.

To our knowledge, such an approach was attempted in Ref.~\cite{Vanneste:2019jnz}, where cross-spectra from \Planck~and WMAP were combined into a single concatenated spectrum (see \autoref{eq: datavec_ohl}). Something similar was also applied in Ref.~\cite{bicepPlanck2015jointAnalysis}, where the \Planck~single-frequency spectra are included as cross-spectra of two data-split maps, treating them as auto-correlations on the diagonal of \autoref{eq: full_hl_mat}. 

Of course, when $\nchs =2$ the cHL is identical to the original oHL, given that it yields only one cross-spectrum. We study this case in \autoref{sec: 2ch}.

We conclude this section by emphasizing that cross-spectra are not the only reason $C_\ell$ can lose positive definiteness. When estimated on a partial sky, even auto-spectra can be negative \cite{hamimeche2008LikelihoodAnalysisCMBTemperature}. The more aggressive the mask, the more likely the estimated auto-spectra will exhibit negative elements. While the offset in the oHL (and in the cHL) helps mitigate this, no such safeguard is included in the HL likelihood defined at the beginning of the section.
Although the partial-sky case is, in principle, distinct from having two anti-correlated fields—since negative spectra in the former arise from the estimation process itself—introducing an offset in the diagonal of \autoref{eq:3chs_full_hl_mat} may still be beneficial, despite the presence of noise bias. We investigate this possibility in the following.

\subsection{The new auto-spectra marginalized Hamimeche \& Lewis} \label{sec: mHL}

We now propose a variant of the HL likelihood, which handles noise bias uncertainty by marginalizing over the auto-spectra, named for brevity the \textit{marginalized HL} (mHL). The procedure is very similar to the one described in \autoref{sec: hl_like}, where, in our case, we start from matrices as \autoref{eq:3chs_full_hl_mat}. However, when computing the likelihood of \autoref{eq:hl_likelihood}, we employ only the ``cross-elements'' of the $X_{\ell}$ data vector defined in \autoref{eq: hl_Xell}:

\begin{equation}
   X_{\ell}^\ncross = \qty(X_\ell^\axb,~ X_\ell^\axc,~ X_\ell^\bxc)\ .
\label{eq: mhl_Xell}
\end{equation}

Correspondingly, the elements of the matrix $M_{\ell\ell^\prime}$ associated with the ``auto-elements'' (e.g. $X_\ell^\axa$) are removed. This approach relies on a simple but often overlooked observation: the HL likelihood has the structure of a Gaussian distribution for the $X_\ell$, with covariance $M_{\ell\ell^\prime}$. Indeed, it can be shown empirically that $X_{\ell} \sim \mathcal{N}\qty(\mu, M_{\ell\ell^\prime})$. Therefore, we can apply the standard marginalization rule for a multivariate Gaussian to obtain a likelihood defined only on $X_{\ell}^\ncross$, without the need to specify $\mu$ \footnote{For a random variable $Y = \qty(Y_1, Y_2, Y_3) \sim \mathcal{N}(\mu, \Sigma)$ with $\mu = \qty(\mu_1, \mu_2, \mu_3)$ and $\Sigma$ a $3\times3$ covariance matrix, the marginal distribution on $Y_2$ is:
\begin{equation}
\begin{aligned}
    \Tilde{Y} &\sim \mathcal{N}\qty(\Tilde{\mu}, \Tilde{\Sigma}) \\
    \Tilde{\mu} & = \qty(\mu_1, \mu_3) \\
    \Tilde{\Sigma} &= \qty(\begin{matrix}
        \Sigma_{11} & \Sigma_{13} \\
        \Sigma_{13} & \Sigma_{33} \\
    \end{matrix})\ .
\end{aligned}
\end{equation}}.
Note that, if $X_{\ell} \sim \mathcal{N}\qty(\mu, M_{\ell\ell^\prime})$ this procedure is fully analogous to performing the integral of the HL likelihood over the auto-elements of $X_\ell$, similarly to what is done in Ref.~\cite{Percival_2006} to obtain the probability of $\widehat{C}_\ell^{\rm TE}$ marginalized over $\widehat{C}_\ell^{\rm TT}$ and $\widehat{C}_\ell^{\rm EE}$. The difference is what is being marginalized: in the mHL we integrate out the $X_{\ell}^{\rm auto}$, treated as random variables, exploiting the fact that the result of that integral is known for a multivariate Gaussian; instead, Ref.~\cite{Percival_2006} marginalizes over the individual $\widehat{C}_{\ell}^\axa$.

Also, we emphasize that the mHL and the HL likelihood share the same assumption of a Wishart-distributed $C_\ell$. In turn, this translates into assuming a Gaussian and statistically isotropic field in pixel-space; thus, any violation of these properties makes both the mHL and the HL likelihood approximations of the true unknown distribution. The only difference is the information retained for parameter estimation: the HL exploits the full information encoded in $X_{\ell}$ (see \autoref{eq: hl_Xell}), while the mHL just the one in $X_{\ell}^\ncross$ (see \autoref{eq: mhl_Xell}).

Hence, a key advantage of this method is its flexibility: it allows one to selectively retain any well-estimated auto-spectrum simply by adjusting the data vector and covariance accordingly. For instance, if the auto-spectrum for the C-channel $C_\ell^\cxc$ is deemed reliable, one could define $X_\ell = \qty(X_\ell^\cxc, X_\ell^\axb, X_\ell^\axc, X_\ell^\bxc)$. This is not as straightforward to do in the cHL framework. Moreover, since the algebra used to construct $X_\ell^\ncross$ is identical to that of the HL likelihood, each element remains unchanged; for instance, $X_\ell^{\axb, \rm mHL} = X_\ell^{\axb, \rm HL}$. This is not the case for the cHL (see \autoref{sec: oHL}).

In principle, this approach avoids defining an offset, or modifying \autoref{eq: g_func}, since it uses the original HL formulation. However, as already mentioned, when dealing with partial-sky data, negative eigenvalues can arise in the HL transformation from the matrix $\mathbf{C}_\ell^{-1/2}\hat{\mathbf{C}}_\ell \mathbf{C}_\ell^{-1/2}$. To mitigate this cut-sky effect and stabilize the eigenvalues, we add an offset only to the diagonal elements of the various matrices involved, as mentioned above for the HL. Specifically, we assume the same offset definition used in the oHL analysis of \cite{mangilli2015LargescaleCMBTemperaturepolarization}. Differently from the oHL (and cHL), we compute this offset for the auto-spectra, e.g. $O_\ell^\axa$, which is typically smaller than its cross counterparts in the cHL, say $O_\ell^\axb$. Moreover, as $\fsky$ goes to $100\%$, $O_\ell^\axa$ goes to zero since no auto-spectrum is negative on the full sky, where the power spectrum estimation does not introduce this artifact. This does not happen for $O_\ell^\axb$: the noise realizations are independent; thus, if the noise is large enough with respect to the signal, the final $C_\ell$ can be negative when the noise anti-correlates between channels.
Furthermore, to ensure complete numerical stability, we use the same change of the $g$ function as for the oHL (see \autoref{eq:g_func_oHL}).

In \autoref{app: offset}, we show the performance of both the HL and mHL likelihoods when this regularization step is omitted and when we use two possible definitions of the offset.

\subsection{Pixel-based likelihood}

Finally, we emphasize once again that none of the likelihoods discussed so far are exact on a cut sky. On the partial sky, $C_\ell$ does not distribute as a Wishart; thus, we must assume otherwise to apply the HL prescription. For this reason, we also compute the Pixel-Based (PB) likelihood \cite{gerbinoLikelihoodMethodsCMB2020}, which is exact under the assumption of Gaussian-distributed pixel power. Since we consider only map simulations obtained by summing CMB and white-noise realizations, we can safely adopt this assumption and compute analytically the pixel covariance matrix. The PB likelihood serves therefore as a reference against which the various approximations are evaluated, HL included. The PB is computed with the publicly available \texttt{parameter\_estimation} algorithm \footnote{\href{https://baltig.infn.it/cosmology_ferrara/lowell-likelihood-analysis/-/tree/master/parameter_estimation}{baltig.infn.it/cosmology\_ferrara/lowell-likelihood-analysis/-/tree/master/parameter\_estimation}.}.

As mentioned above, the three channels we study in this work can also be seen as three data-splits of a hypothetical mission. Thus, we compute the PB on the full-mission dataset, after combining the three maps from the channels. In this sense, the PB will contain the full information, which we try to recover with the multi-field HL-based likelihoods. 

\section{Methodology} \label{sec: methodology}

\subsection{Our setup} \label{sec: setup}

In this work, we assess the robustness of HL-based likelihoods against misestimation of the noise bias. We consider either \clee or \clbb with $\nchs = 3$ channels, within the context of both the \lcdm~and \lcdmr~models, to estimate $\tau$ and $\rpivot$, respectively (see \autoref{sec: 2ch} for the $\nchs = 2$ case). Specifically, we define three test cases:

\begin{table}[t]
\centering
\caption{\label{tab:fiducial_params}
Fiducial values for the cosmological parameters. In parenthesis we report the value of the tensor-to-scalar ratio in the tensor-detection case.}
\begin{tabular}{lc}
\toprule
\textrm{Parameter}& \textrm{Fiducial value} \\
\midrule
$\As 10^{-9}$ & $2.12$ \\
$\ns$ & $0.9651$ \\ 
$\omegab h^2$ & $0.02237$ \\
$\omegac h^2$ & $0.1201$ \\
$\hnot$ & $67.32~\text{km/s/Mpc}$ \\
$\tau$ & $0.06$ \\
$\rpivot$ & $0$ ($0.01$) \\
\midrule
$\mnu$ & $0.06$ \\
\bottomrule
\end{tabular}
\end{table}

\begin{itemize}
    \item Signal-dominated case: using E-modes to estimate the optical depth $\tau$ in a regime where the signal dominates over the noise. In particular, we assume a white-noise level of $N^{\rm E} = 80~\mu$K-arcmin for each of the three E-mode channels ($A$, $B$, and $C$).
    \item Noise-dominated case: using B-modes under the null-signal-hypothesis of $\rpivot = 0$. Here, we instead assume $N^{\rm B} = 10~\mu$K-arcmin for all channels.
    \item Tensor-detection case: again using B-modes, but now with $\rpivot = 0.01$, representing an intermediate regime between the previous two. Indeed, given that we assume the same noise level of the noise-dominated case, we allow for a detectable $\rpivot$ without being completely signal-dominated.
\end{itemize}

For both the noise-dominated and tensor-detection cases, we assume the presence of the so-called lensing B-mode contribution due to the deflection of E-modes caused by mass distributions. On top of this, maps are smoothed with a cosine beam to limit their harmonic content without sharp cuts \cite{aghanimPlanck2018Results2020a, benabed2009teasing}.

The fiducial values of the cosmological parameters are chosen to be consistent with current constraints \cite{planckcollaboration2020Planck2018ResultsVI, rosenberg2022CMBPowerSpectraCosmological, Tristram:2023haj, collaboration2021BICEPKeckXIIIImproved, galloniUpdatedConstraintsAmplitude2022} and are summarized in \autoref{tab:fiducial_params}. Of course, the tensor-detection case is an exception, as we set $\rpivot = 0.01$ instead. All theoretical CMB angular power spectra are computed with \CAMB \cite{lewisEfficientComputationCosmic2000}.

\subsection{Simulations}

For each fiducial power spectrum, we generate $N = 10000$ simulations using the \texttt{healpy} package. Maps are created at $N_{\rm side}= 16$, corresponding to an angular resolution of approximately $220'$ and a total of 3072 equal-area pixels over the full sky. Harmonic information is extracted using a Quadratic Maximum Likelihood (QML) algorithm. Similarly to the PB likelihood, the QML requires the computation of the pixel covariance matrix, understood as the sum of a signal contribution from CMB and a diagonal noise contribution (isotropic white noise). While not required for full-sky coverage, QML ensures consistency with our cut-sky analysis. In particular, we adopt the \texttt{pse\_qml} \footnote{\href{https://baltig.infn.it/cosmology_ferrara/lowell-likelihood-analysis/-/tree/master/pse_qml}{baltig.infn.it/cosmology\_ferrara/lowell-likelihood-analysis/-/tree/master/pse\_qml}.} algorithm.

The first $N_{\rm sims} = 9000$ simulations are used to estimate the harmonic covariance matrix, the fiducial spectra, and eventually the offset (see \autoref{eq:hl_likelihood}). The remaining $N_{\rm data} = 1000$ form the validation set. To avoid pixelization effects, we set the maximum multipole to $\ell_{\rm max} = 2 \times$ $N_{\rm side} = 32$, in line with typical large-scale CMB analyses that use $\ell_{\rm max} \simeq 30$.

\begin{figure}
    \centering
    \includegraphics[width=\linewidth]{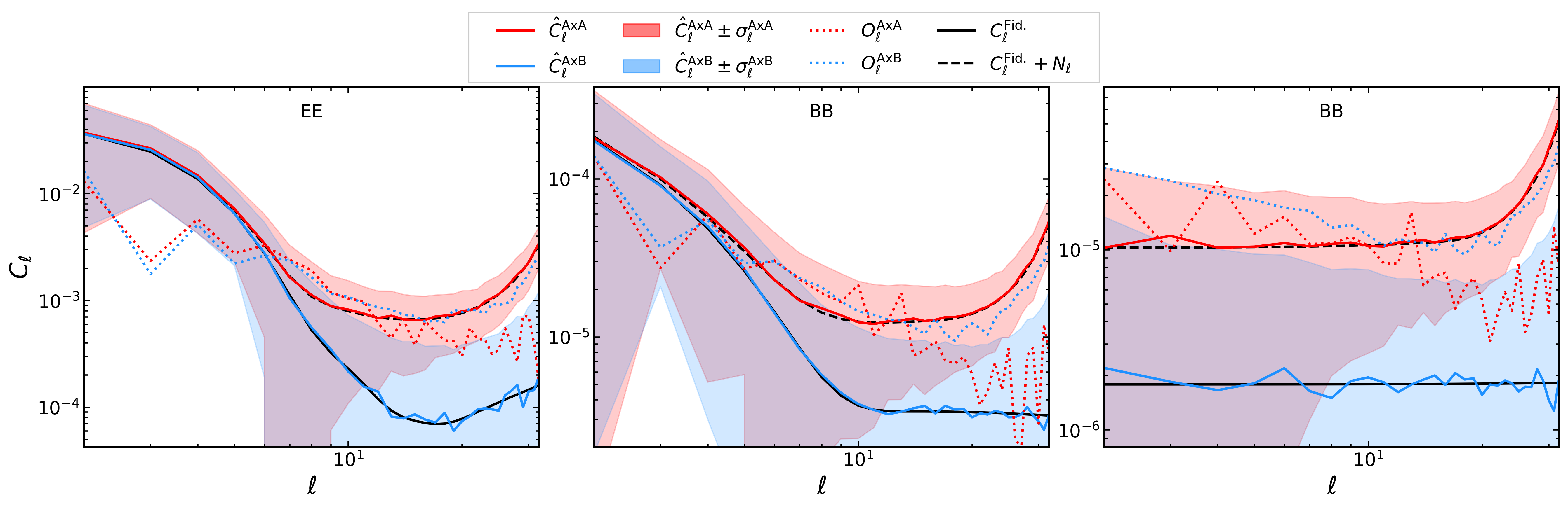}
    \caption{Auto- (red) and cross-spectra (blue) for the signal-dominated, tensor-detection, and noise-dominated cases, respectively. Solid lines indicate the mean over the $N_{\rm data}$ simulations, while shaded regions refer to the $1\sigma$ scatter around the mean and dotted ones to the corresponding offset $O_\ell$. As a reference, we also show the fiducial power spectrum with and without noise bias with black solid and dashed lines.}
    \label{fig: spectra_mosaic}
\end{figure}

As mentioned above, each simulation is characterized by some level of white-noise (see \autoref{sec: setup}). To mimic misestimation of these noise levels, we generate additional simulations in which the first $N_{\rm sims}$ simulations use incorrect noise amplitudes. Specifically, we apply a $\pm 20\%$ scaling to the true noise at the map level, both over- and underestimating it. We denote the estimated noise power spectrum as $\widehat{N}_\ell$ and the true one as $N^{\rm True}_\ell$. This 20\% error is a deliberately pessimistic assumption \footnote{Note that a $20\%$ misestimation at map-level brings to a $\simeq40\%$ mismatch in harmonic space.}. Note that in this work we focus on the likelihood characterization of the CMB sky. Thus, this mismatch is never introduced in the power spectrum estimation step of our analysis. In a real-life scenario, these two would combine; however, here we isolate and study solely the likelihood effect.

To evaluate stability with respect to the fiducial cosmological value assumed, we also vary the fiducial value of $\rpivot$ in the noise-dominated and tensor-detection cases, swapping them between 0 and 0.01, and consider an additional scenario with $\rpivot = 0.02$. For the signal-dominated case, we explore values $\tau = 0.03$ and $\tau = 0.09$ to test both under- and overestimation. Of course, this choice affects just the first $N_{\rm sims}$ simulations to mimic once again a wrong estimate. These values of the fiducial are deliberately extreme tests, as in real data analyses, strongly biased fiducials would be obvious and can be corrected through iterative re-analysis. As such, we relegate the results of these tests to \autoref{app: wrongfid}.

We also repeat all analyses under partial-sky conditions. Although the full-sky case highlights the importance of proper likelihood usage under ideal conditions, we are also interested in more realistic scenarios with incomplete sky coverage. We use the publicly available HFI masks that retain 40\%, 60\%, and 80\% of the sky, respectively \footnote{\href{https://pla.esac.esa.int/\#maps}{pla.esac.esa.int/\#maps}.}. Each simulation is masked, and its power spectrum is estimated using QML. This allows us to examine any difference in likelihood behavior caused by the cut-sky. To isolate the effects of mismatches in the likelihood, we always assume the correct noise and fiducial values when estimating the power spectra. Studying the effect of incorrect assumptions during power spectrum estimation lies beyond the scope of this work and has been addressed elsewhere in the literature \cite{Tegmark:1994we, tegmark1997HowMeasureCMBpower, Bond:1998zw, Efstathiou:2003dj}.

Before discussing how we recover the posterior of each case, we show in \autoref{fig: spectra_mosaic} an example of the spectra we are using for the signal-dominated, tensor-detection, and noise-dominated cases, respectively. In particular, we choose $\axa$ and $\axb$ as representatives of auto- and cross-correlations, plotting their recovered spectra with $\fsky = 40\%$ \footnote{We do not show the full-sky case as it is trivial.}. In solid lines, we show the mean of the $N_{\rm data}$ simulations, the shaded regions indicate the one-sigma scatter, and the dotted lines the offsets. In addition, we show the fiducial spectrum with and without the noise bias as a reference. Here, the effect of the beam function mentioned above is clearly visible as a degradation of the noise bias towards small scales. This plot shows that our QML estimation is unbiased. Also, one can appreciate that the offsets of $\axa$ are comparable with the ones of $\axb$, given the large number of negative spectra found assuming $\fsky = 40\%$. For the sake of brevity, we do not show this here, but while $\fsky$ increases, the two offsets decrease and depart from each other, with $O_\ell^\axa$ tending to zero faster than $O_\ell^\axb$ (see definition in \autoref{sec: oHL}).

\subsection{Posterior evaluation}

To validate all these HL-based likelihoods, we aim to estimate $\tau$ from E-modes and $\rpivot$ from B-modes, separately and with all other cosmological parameters held fixed (see \autoref{tab:fiducial_params}). Thus, we evaluate the likelihood over a fixed grid of values on the considered parameter.

We define a uniform grid for $\tau$ in the range $[0.03, 0.09]$ using 960 steps, and for $\rpivot$ in the range $[0.0, 0.02]$ using 800 steps. This is equivalent to assuming a flat prior over the specified ranges \cite{Pagano_2020}.

Using this setup, we compute the posterior for each of the $N_{\rm data}$ simulations. Doing the product over these posteriors yields a single mean posterior, which allows us to assess whether any bias has been introduced during the analysis pipeline.

The code developed and used to perform this analysis is publicly available as a \texttt{Python} package named CMB Harmonic Analysis with Robust Multi-field Likelihood (\href{https://github.com/ggalloni/CHARM-Like}{\texttt{CHARM-Like}}). Providing a set of precomputed angular power spectra for some channels, \texttt{CHARM-Like} allows us to quickly compute the posteriors of HL, mHL and cHL, together with all the single field HL.


\section{Results} \label{sec: results}

\subsection{Full-sky}\label{sec: res_fullsky}

We start from the full-sky case, where every component of the analysis is perfectly characterized. In particular, the HL result is exact, and no mode coupling arises from masking effects.
\begin{figure}
    \centering
    \includegraphics[trim={0 9cm 0 0},clip, width=\linewidth]{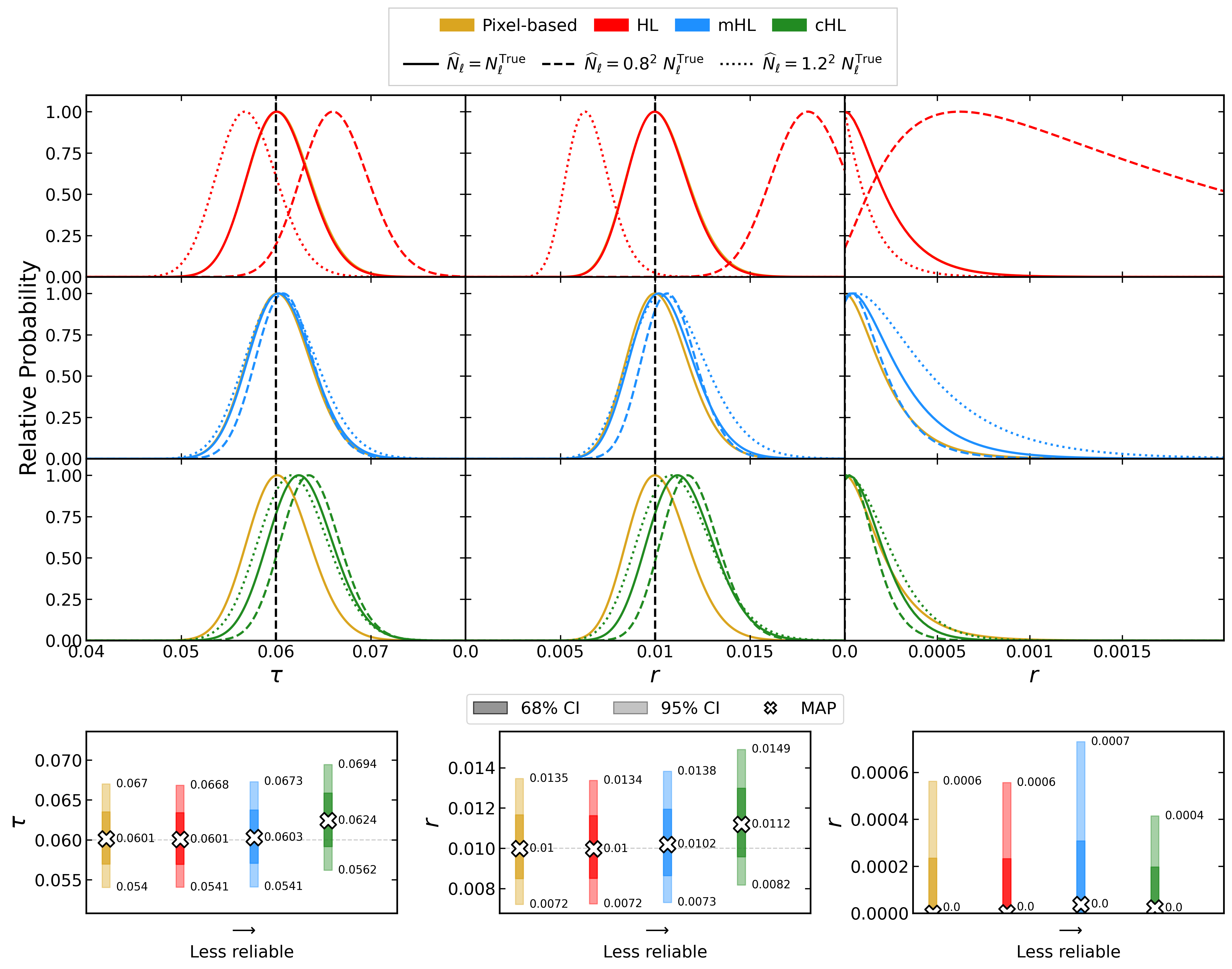}
    \caption{Posterior distributions from pixel-based (gold), HL (red), mHL (blue), and cHL (green) likelihoods under full-sky assumptions. Dashed (dotted) curves indicate under- (over-) estimation of the noise bias. Vertical black lines denote the true parameter values.}
    \label{fig: mismatch_BB_oHL_mHL}
\end{figure}
\autoref{fig: mismatch_BB_oHL_mHL} summarizes the results for the HL, mHL, and cHL likelihoods, including cases where the noise bias is misestimated. Each row (color-coded) corresponds to a different likelihood, while each column refers to the signal-dominated, tensor-detection, and noise-dominated cases, respectively. Dashed lines represent an underestimation of the noise bias, and dotted lines represent an overestimation.

Starting with the HL, all posteriors are unbiased when the correct noise bias is used. In fact, they appear exactly identical to the PB results. The same applies to the single-field HL, which also overlaps perfectly with the PB, and so is not shown here for clarity.

However, when a mismatch is introduced, a bias appears \footnote{Here, we do not show the PB and the single-field HL with noise bias misestimation. In fact, those two serve for comparison with the other HL-based likelihoods, and it is known that they are quite sensitive to such a mismatch.}. As expected, the bias is less severe when the signal dominates (left column of \autoref{fig: mismatch_BB_oHL_mHL}). Both mHL and cHL mitigate this bias to varying degrees in all cases: in the signal-dominated case the mHL shows a smaller dispersion with respect to the cHL, thus performing better. In the tensor-detection case, they perform similarly, while in the noise-dominated case the cHL behaves better. Indeed, even though auto-correlations are marginalized over, the incorrect noise still enters the HL transformation algebra.

The mHL generally yields more conservative results. This stems from the marginalization, effectively discarding some information captured by the HL. However, this degradation is significant only in the noise-dominated case (and affects marginally the tensor-detection case), suggesting a dependence on the signal-to-noise ratio.

\begin{figure}
    \centering
    \includegraphics[width=.5\linewidth]{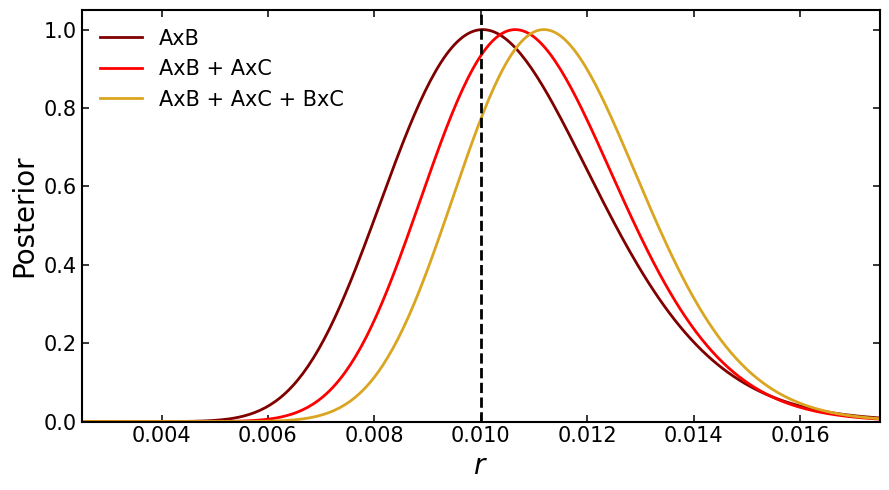}
    \caption{Dependence of cHL on the number of cross-spectra used. Results are shown for one, two, and three spectra for the tensor-detection case in full-sky.}
    \label{fig: number_field_oHL}
\end{figure}

A striking feature in \autoref{fig: mismatch_BB_oHL_mHL} is the bias observed in the cHL even when the correct noise bias is used. This originates from applying the HL transformation independently to each cross-spectrum (see \autoref{eq: datavec_ohl}). This bias increases with the number of cross-spectra used; this is shown in \autoref{fig: number_field_oHL} for the tensor-detection case.

With a single cross-spectrum, as in \cite{tristram2022ImprovedLimitsTensortoscalarratio}, the bias is null, but it becomes more pronounced when multiple channels are included.

\subsection{Cut-sky} \label{sec: res_cutsky}

Having validated the likelihoods in an ideal full-sky context, we now turn to a more realistic scenario with partial sky coverage. In this case, even the HL likelihood is no longer exact, and the PB approach must serve as our benchmark.

For brevity, we present results for $\fsky = 40\%$, the most extreme case considered. Additional results for $\fsky = 60\%$ are provided in \autoref{app: offset}, while the $\fsky = 80\%$ case closely resembles the full-sky scenario and is therefore omitted.

\begin{figure}
    \centering
    \includegraphics[width=\linewidth]{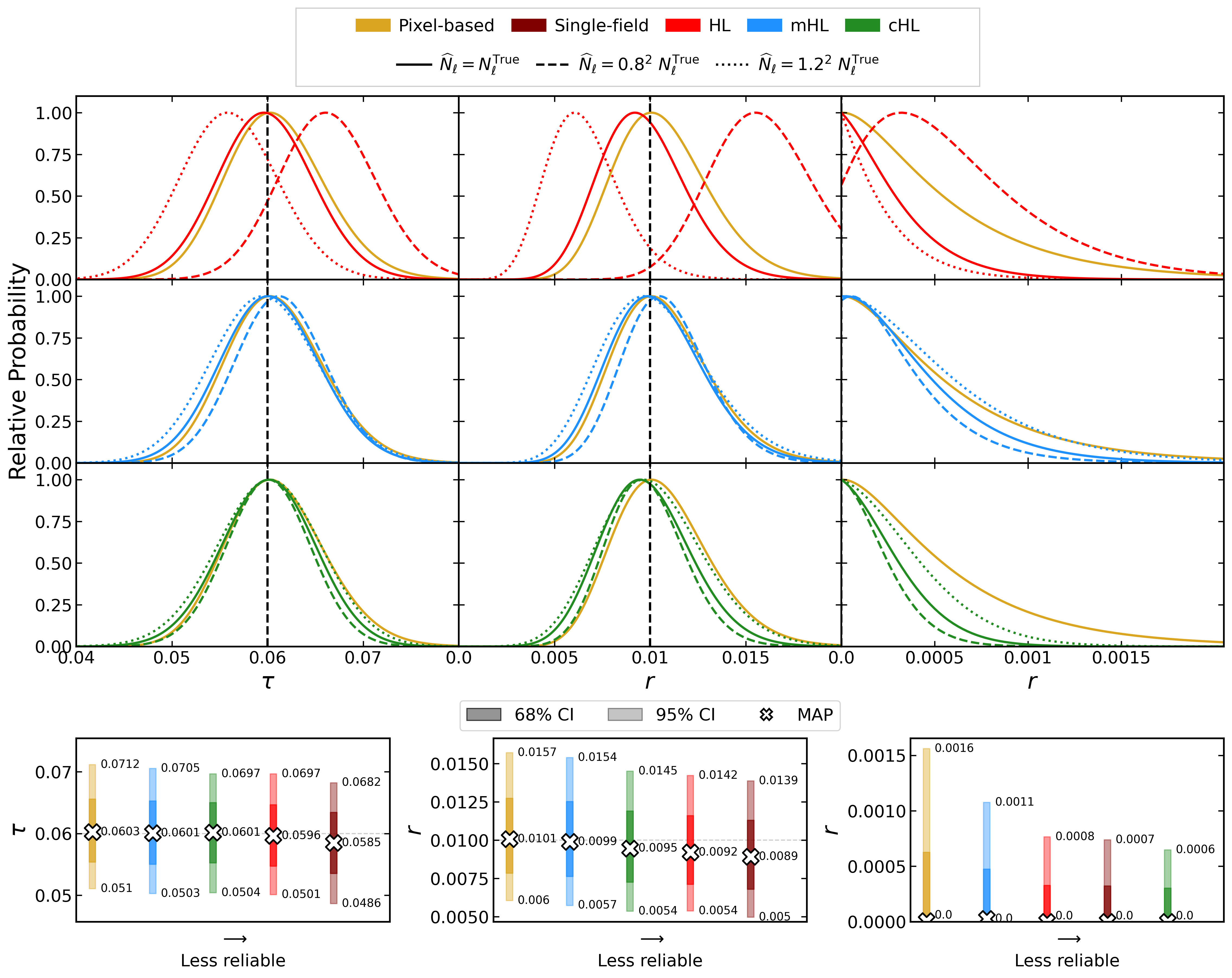}
    \caption{Posterior distributions from pixel-based (gold), HL (red), mHL (blue), and cHL (green) likelihoods for $\fsky = 40\%$. The columns correspond to signal-dominated, tensor-detection, and noise-dominated cases. In the first three rows, dashed (dotted) curves correspond to under- (over-) estimated noise bias. Vertical black lines mark the true parameter values ($\tau = 0.06$, $\rpivot=0.01$ and $\rpivot=0$). The bottom row provides a quantitative comparison of the solid-line posteriors above, together with the single-field HL (maroon). Crosses show MAP estimates, while shaded bars indicate 68\% and 95\% credible intervals.}
    \label{fig: offset_fsky40_mismatch_BB_oHL_mHL}
\end{figure}

\autoref{fig: offset_fsky40_mismatch_BB_oHL_mHL} shows the results under noise bias mismatch for $\fsky = 40\%$ and summarizes the main results of this work. The first three rows, analogous to \autoref{fig: mismatch_BB_oHL_mHL}, clearly show broader posteriors, a direct consequence of information loss due to the mask \footnote{Here, we do not plot the single-field HL posteriors for visualization purposes. Instead, we relegate it to the bottom panel, which is described below.}.

Focusing on the solid lines (i.e., with correct noise bias) and leaving aside the noise-dominated case for a moment, all likelihoods perform reasonably well in recovering the pixel-based results. Nonetheless, the mHL stands out as the best-performing approximation, consistently matching the PB across all cases, without showing the same biases affecting the others in the tensor-detection case, for example. In the noise-dominated case, the better accuracy of the mHL is particularly evident when comparing it to the others. We emphasize the crucial role played by the offset, as further discussed in \autoref{app: offset}, where we study the effect of removing it in the computation of both the HL and the mHL. In addition, we test in the same appendix two different offset prescriptions \cite{mangilli2015LargescaleCMBTemperaturepolarization, tristram2022ImprovedLimitsTensortoscalarratio}. \autoref{app: offset} highlights another strength of the mHL: its robustness to the arbitrary choice of offset, owing to the fact that it marginalizes over the auto-spectra where the offset is introduced.

The HL shows a strong signal-to-noise dependence. While $\tau$ is well recovered in the signal-dominated case, it significantly underestimates uncertainty on $\rpivot$ in the noise-dominated case. In contrast, mHL remains robust, even improving over its full-sky performance in the noise-dominated case. This can be pictured as the cut-sky introducing an additional scatter in the estimated power spectra—an extra ``noise contribution'' not accounted for, which the mHL marginalizes over.

cHL shows a systematic shift to the left due to sky masking, increasing in magnitude as the noise becomes more important \footnote{Not shown here for brevity, but reducing $\fsky$ from 100\% to 40\% progressively shifts the cHL leftward with respect to the full-sky result.}. Thus, in the signal-dominated case, the leftward bias due to the mask is balancing the rightward bias affecting the cHL in full-sky (see \autoref{fig: mismatch_BB_oHL_mHL}). In other words, this is serendipitous as shown in \autoref{app: offset} for $\fsky=60\%$.
Similarly to the HL, the cHL strongly underestimates the uncertainty on $\rpivot$ in the noise-dominated case. Indeed, similarly to the HL, the posterior approaches $\rpivot = 0$ with a negative derivative, indicating that the actual peak of the posterior would be in the non-physical region $\rpivot < 0$ (a similar leftward bias is seen also in the tensor-detection case).

As for the impact of a mismatch in the noise bias, the results align with those in \autoref{fig: mismatch_BB_oHL_mHL}, suggesting that the effects of noise mismatch and partial sky coverage interact in a trivial way. The only notable difference is that the noise mismatches lead here to smaller biases. This is simply understood as the mask itself is producing an extra scatter, making the actual noise bias less important than before.

Finally, the bottom row of \autoref{fig: offset_fsky40_mismatch_BB_oHL_mHL} presents a quantitative comparison between the posteriors with correctly estimated noise bias (solid lines in the upper panel of the same figure). Here, we also show the results of the single-field HL, of which we do not plot the posterior for visualization purposes. The approximations are ranked by their ability to recover the PB results. While this is qualitatively evident, we compute the Kullback–Leibler (KL) divergence between the PB and each approximation to assess quantitatively this correspondence. Thus, the KL divergence allows us to rank them accordingly. In particular, we show the Maximum A Posteriori (MAP) estimate and the 95\% credible interval for each case.

This ranking confirms that mHL offers the best agreement with PB while maintaining robustness to noise bias mismatches. The worst performance occurs in the noise-dominated case, where even the mHL underestimates the 95\% upper bound by $\simeq 28.8\%$. The HL and cHL perform significantly worse, with underestimations of approximately $\simeq 49.5\%$ and $\simeq 57.1\%$, respectively.

It is also worth noting that in all noise mismatch cases, the mHL provides the best KL divergence with respect to the true posterior. The only exception is the cHL in the signal-dominated case, which scores slightly better than the mHL. This means that, despite a possible misestimation of the noise bias, mHL still gives the most accurate and robust posterior among the ones studied in almost all the cases considered.

In addition, we note that the single-field HL tends to align with the HL. This suggests that applying the HL likelihood on the full-mission dataset, or on multiple data splits, is nearly equivalent when one employs all fields. This emphasizes the strength of the mHL, which allows for improvements in accuracy with respect to the PB, while offering robustness against an eventual mismatch of the noise bias. Thus, this suggests that a multi-field mHL likelihood should be preferable to a single-field HL for CMB analyses.

To conclude this section, we repeat the analysis while adopting the familiar, though notoriously incorrect, assumption that the $C_\ell$ distribution is Gaussian. Even though this is a rougher approximation with respect to the HL on large scales \cite{gerbinoLikelihoodMethodsCMB2020}, it is interesting to see how that performs. In \autoref{app: gaussian}, we show that this approximation brings distortions of the posterior, as expected.

\subsection{Using only two channels} \label{sec: 2ch}

In the previous section, we focused on a multi-field case with three ``channels'', or data splits, observing the same CMB signal with different noise realizations. This setup gives us three cross-spectra to build the mHL and cHL approximations.

In \autoref{sec: res_fullsky}, we mentioned a notable feature of the cHL approximation that depends on the number of cross-spectra used. While we do not explore this effect further, it is useful to examine the two-channel case more closely. This setup is common in the literature and is the only multi-field configuration with a single cross-spectrum \cite{Pagano_2020, tristram2022ImprovedLimitsTensortoscalarratio}.

We repeated the analysis from \autoref{sec: methodology} using only channels A and B. In this case, the cHL likelihood definition is identical to the oHL, so we will use the latter name to remark this fact.

\begin{figure}
    \centering
    \includegraphics[width=\linewidth]{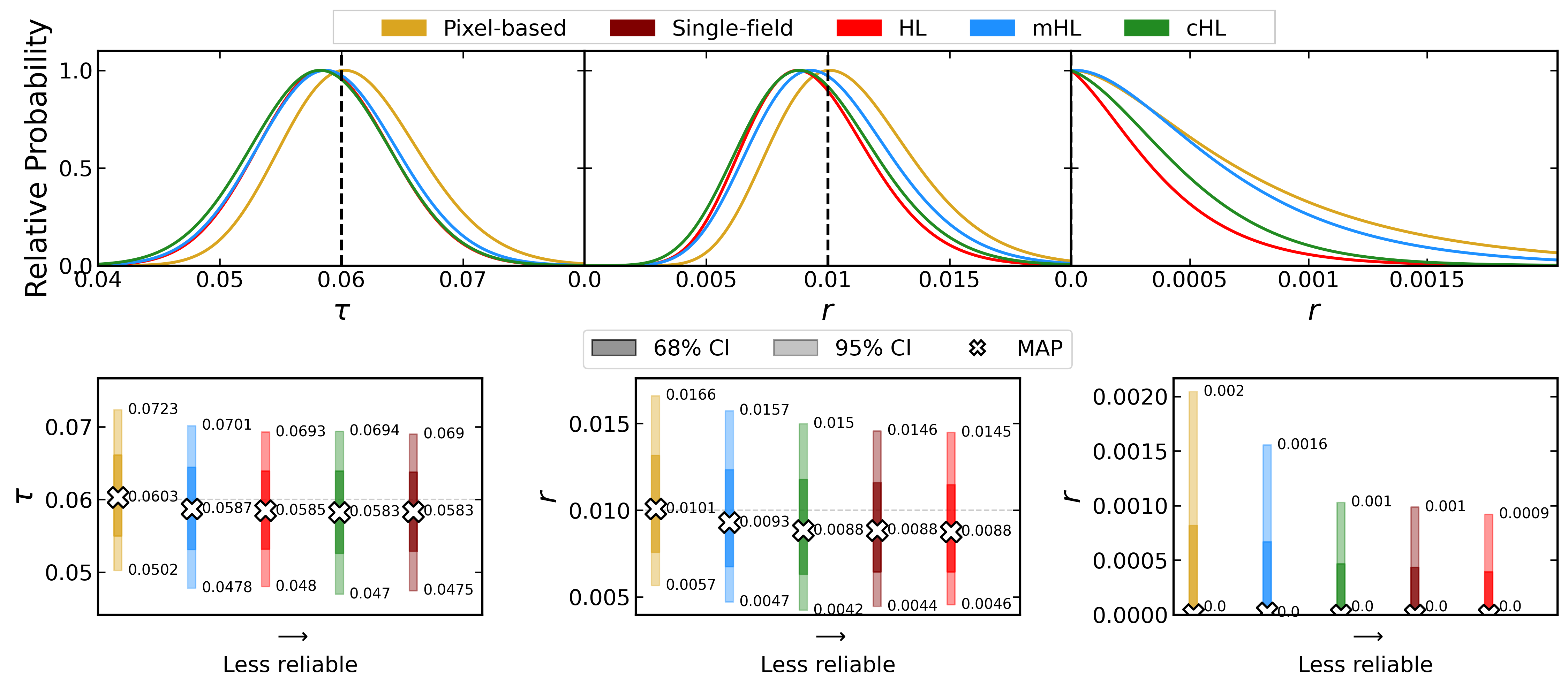}
    \caption{Comparison between pixel-based (gold), HL (red), mHL (blue), and cHL (green) likelihoods with $\fsky = 40\%$ using only channels A and B. In this case, cHL and oHL are identical. The columns correspond to signal-dominated, tensor-detection, and noise-dominated cases. The black vertical lines represent the true values of the parameters ($\tau = 0.06$, $\rpivot=0.01$ and $\rpivot=0$). The bottom row provides a quantitative comparison of the posteriors above, together with the single-field HL (maroon). Crosses show MAP estimates, while shaded bars indicate 68\% and 95\% credible intervals.}
    \label{fig: fsky40_BB_oHL_mHL_2ch}
\end{figure}

\autoref{fig: fsky40_BB_oHL_mHL_2ch} shows the results for $\fsky=40\%$. There are two notable differences compared to the three-channel case. First, in the signal-dominated and tensor-detection cases, the HL and mHL do not match the PB results perfectly. Still, all three approximations—HL, mHL, and oHL—give similar posteriors, broadly consistent with the true one. The mHL performs better overall.

In the noise-dominated case, the HL and oHL underestimate the uncertainty on $\rpivot$, as seen in \autoref{sec: results}. In contrast, the mHL performs better than the others. In particular, they all underestimate the 95\% upper bound: the mHL is off by $\simeq 23.9\%$, the oHL by $\simeq 49.7\%$, and the HL by $\simeq 55\%$.

Finally, the single-field HL tends to align with the HL also in this case. This suggests once again that it is equivalent to apply the HL approach to one field or to multiple ones. Let us emphasize that here the full-mission dataset used for this single-field HL and the PB is obtained by combining solely channels A and B, to be consistent with the other approximations.

These results, however, are not directly comparable with the literature \cite{tristram2022ImprovedLimitsTensortoscalarratio}. Indeed, here we use a different prescription for the offset. Similarly to \autoref{app: offset}, we repeat this two-channel analysis using LoLLiPoP offset \cite{tristram2022ImprovedLimitsTensortoscalarratio}. For the sake of brevity, we do not show the results here. However, in this context, oHL and mHL provide very similar results in the noise-dominated case, with the oHL performing slightly better. On one side, this supports the use of oHL in single-cross-spectrum analyses of noise-dominated scenarios (e.g., \cite{tristram2022ImprovedLimitsTensortoscalarratio}), and shows that the mHL is a competitive alternative. On the other hand, this emphasizes once more the importance of the choice of the offset.


\section{Conclusions} \label{sec: conclusions}

In this work, we investigated the performance of several likelihood approximations based on the Hamimeche-Lewis (HL) prescription \cite{hamimeche2008LikelihoodAnalysisCMBTemperature}, focusing in particular on two extensions: the newly introduced \textit{marginalized HL} (mHL) and the cross-spectra HL (cHL)--an extension of the offset HL (oHL) likelihood   \cite{mangilli2015LargescaleCMBTemperaturepolarization}. We studied the impact of realistic complications, such as partial sky coverage, mismatches in noise bias estimation, and incorrect fiducial assumptions. Our analysis was carried out in a multi-field context, where three observational channels, or data splits, of CMB polarization were jointly used for parameter estimation. Given that the distribution of data is more easily approximated by the HL likelihood on intermediate and small scales \cite{hamimeche2008LikelihoodAnalysisCMBTemperature}, thanks to the central limit theorem, we focused on the large-scale part of CMB, where the harmonic approximation represents a bigger challenge.

We began by validating each method in the idealized full-sky case (\autoref{fig: mismatch_BB_oHL_mHL}). As expected, the HL delivered unbiased and precise results when the noise bias was correctly modeled, but suffered from significant biases when it was not. Both the mHL and cHL mitigated these biases, with the mHL providing more conservative but robust constraints. We also identified a bias in the cHL when combining multiple cross-spectra, linked to the number of spectra considered: the more one uses, the worse the bias becomes (see \autoref{sec: res_fullsky}). In \autoref{app: wrongfid}, we showed that the HL is particularly stable under fiducial mismodeling as expected from Ref.~\cite{hamimeche2008LikelihoodAnalysisCMBTemperature}. The mHL displayed similar robustness, while the cHL exhibited stronger sensitivity due to its reliance on the fiducial spectrum through the offset.

In the cut-sky scenario, where even the HL becomes approximate, we benchmarked all methods against the pixel-based (PB) likelihood \cite{gerbinoLikelihoodMethodsCMB2020}, computed using the ``full-mission'' dataset obtained by combining the three data splits. The mHL showed excellent agreement with the PB posterior, benefiting from its marginalization over auto-spectra, which helps absorb inaccuracies from noise modeling and sky cuts (\autoref{fig: offset_fsky40_mismatch_BB_oHL_mHL}). Both the HL and cHL tended to underestimate uncertainties, particularly in the noise-dominated case (see \autoref{sec: methodology}). In this context, we also used an offset for the HL and mHL, and we showed in \autoref{app: offset} that the mHL remains stable under different offset prescriptions, thanks to the marginalization over auto-correlations.

To quantitatively compare the methods, we used the Kullback-Leibler divergence as a ranking metric. This confirmed the superior performance of the mHL in recovering the PB posterior across all tested scenarios. Even in the challenging noise-dominated regime, the mHL maintained acceptable accuracy, significantly outperforming both the HL and the cHL.

Using the same full-mission dataset, we also computed a ``single-field HL'' likelihood, corresponding to a single-field case where all data splits are combined into one map. Comparing results in \autoref{fig: offset_fsky40_mismatch_BB_oHL_mHL}, we found that the multi-field mHL approach performs significantly better than the single-field HL, suggesting that a multi-field treatment should be preferred for CMB data analysis. It not only improves accuracy but also enhances robustness against misestimations of the noise bias.

Overall, the marginalized HL emerges as a promising alternative to both the HL and the cHL. It is particularly well-suited for scenarios with imperfect noise knowledge and limited sky coverage, offering a strong balance between robustness and precision.

In \autoref{sec: 2ch}, we tested these methods in the two-channel case, a configuration often used in the literature \cite{Pagano_2020, tristram2022ImprovedLimitsTensortoscalarratio}, which yields a single cross-spectrum. This is particularly interesting because the cHL definition (see \autoref{sec: cHL}) is identical to the oHL one, allowing a comparison with Ref.~\cite{mangilli2015LargescaleCMBTemperaturepolarization} and \cite{tristram2022ImprovedLimitsTensortoscalarratio}. The mHL and the oHL provide very similar results in this case, with slight differences brought by the specific implementation of the offset (see also \autoref{app: offset}). This supports the use of oHL in single-cross-spectrum analyses of noise-dominated scenarios (e.g., \cite{tristram2022ImprovedLimitsTensortoscalarratio}), and shows that the mHL is a viable and robust alternative.

As shown in \cite{hamimeche2008LikelihoodAnalysisCMBTemperature} for the full-sky case, the distribution of the angular power spectrum estimator built from cross-correlations of many maps tends toward a Wishart distribution as the number of maps increases. Specifically, this number scales approximately as $\nchs \gg 1 + N_\ell / \qty(C_\ell + N_\ell)$, depending on the signal-to-noise ratio. When noise is negligible, a single cross-spectrum may be sufficient, while in noise-dominated regimes, many more maps are needed, with $\nchs \gg 2$. Since the HL transformation is designed for Wishart-distributed variables, working closer to this regime is desirable. In this context, both the HL and the mHL likelihoods follow a similar behavior: in the signal-dominated case, they recover the exact solution accurately, but their performance degrades progressively toward the noise-dominated case. Additionally, using $\nchs=2$ rather than $\nchs=3$ yields worse results overall, as expected from the logic presented in \cite{hamimeche2008LikelihoodAnalysisCMBTemperature}. The same does not apply to the cHL, which is more and more biased as the number of cross-spectra increases (see \autoref{app: offset}). Although the mHL currently provides the most accurate approximation to the exact solution among the tested methods, these observations suggest that increasing the number of available maps could further improve performance.

Finally, in \autoref{app: gaussian}, we tested a Gaussian likelihood approximation \cite{gerbinoLikelihoodMethodsCMB2020}, often used for its simplicity. As expected, this assumption fails to capture the non-Gaussian features of the true $C_\ell$ distribution, leading to various distortions of the posteriors (see the appendix for more details). 

Before concluding, it is also important to emphasize that the analysis presented here is applicable to any set of fields, not only data splits of a single observable -- e.g. E-modes in the signal-dominated case. Thus, instead of $\qty{E_{\rm chA}, E_{\rm chB}, E_{\rm chC}}$, one could consider $\qty{T_{\rm chA}, T_{\rm chB}, T_{\rm chC}, E_{\rm chA}, E_{\rm chB}, E_{\rm chC}}$, or any other combination. The only rule to correctly apply the mHL prescription is to first build the full matrix as \autoref{eq: full_hl_mat} starting from the spherical harmonic coefficients of the desired fields. Then, one should apply the HL transformation and marginalize on combinations considered unreliable.

We also emphasize that our analysis assumes an idealized scenario with uncorrelated white noise. In practice, the noise structure can be more complex—particularly in the presence of foregrounds and systematics—which may introduce non-trivial correlations. In this context, the $\ell$–$\ell^\prime$ independence is violated due to the mask; however, we already showed in \autoref{sec: res_cutsky} that the mHL handles this deviation well, yielding accurate and robust results. However, one could also have $m-m^\prime$ couplings and/or $m$-dependence of the covariance matrix, breaking statistical isotropy. Depending on how strong these deviations are, the HL likelihood could cease to be a good approximation, even in the full-sky case. A detailed study of such effects is beyond the scope of this work and is left for future exploration.

Several directions for future work are now open. A more detailed study of the offset computation would be valuable. Although the mHL appears robust against different offset definitions, a systematic investigation is needed to fully assess its stability, together with the dependence of other HL-based likelihoods. It would also be useful to explore the behavior of these likelihoods when increasing the number of fields. Moving from three to four channels is straightforward in principle, but studying the effects systematically could provide further insight. In addition, applying the mHL approach to real datasets could test its practical advantages over current analysis pipelines. Finally, while we focused here on large angular scales, where statistical uncertainties are significant, the same marginalized approach can be applied to smaller scales, e.g. in the context of ground-based experiments such as BICEP/Keck array~\cite{Ade_2014}, Simons Observatory~\cite{Ade_2019}, CMB-S4~\cite{abazajian2016cmbs4}.


\acknowledgments

The authors thank Loris P.L. Colombo, Matthieu Tristram, and W. L. Kimmy Wu for their valuable comments and discussions. Some of the results in this paper have been derived using the following packages: \CAMB~\cite{lewisEfficientComputationCosmic2000}, \texttt{healpy} \cite{gorskiHEALPixFrameworkHighResolution2005}, \texttt{Matplotlib} \cite{hunterMatplotlib2DGraphics2007}, \texttt{SciPy} \cite{virtanen2020SciPyFundamentalAlgorithmsscientific}, \texttt{Numba} \cite{lam2015NumbaLLVMbasedPythonJIT} and \texttt{NumPy} \cite{harrisArrayProgrammingNumPy2020}. We acknowledge the financial support from the INFN InDark initiative and from the COSMOS network through the ASI (Italian Space Agency) Grants 2016-24-H.0 and 2016-24-H.1-2018. M.G. and M.L. are funded by the European Union (ERC, RELiCS, project number 101116027). M.G. acknowledges support from the PRIN (Progetti di ricerca di Rilevante Interesse Nazionale) number 2022WJ9J33. This work was supported by the MUR PRIN2022 Project “BROWSEPOL: Beyond standaRd mOdel With coSmic microwavE background POLarization”-2022EJNZ53 financed by the European Union - Next Generation EU.

\begin{appendix} 

\section{Assuming a wrong fiducial model}\label{app: wrongfid}

\begin{figure} 
    \centering 
    \includegraphics[width=\linewidth]{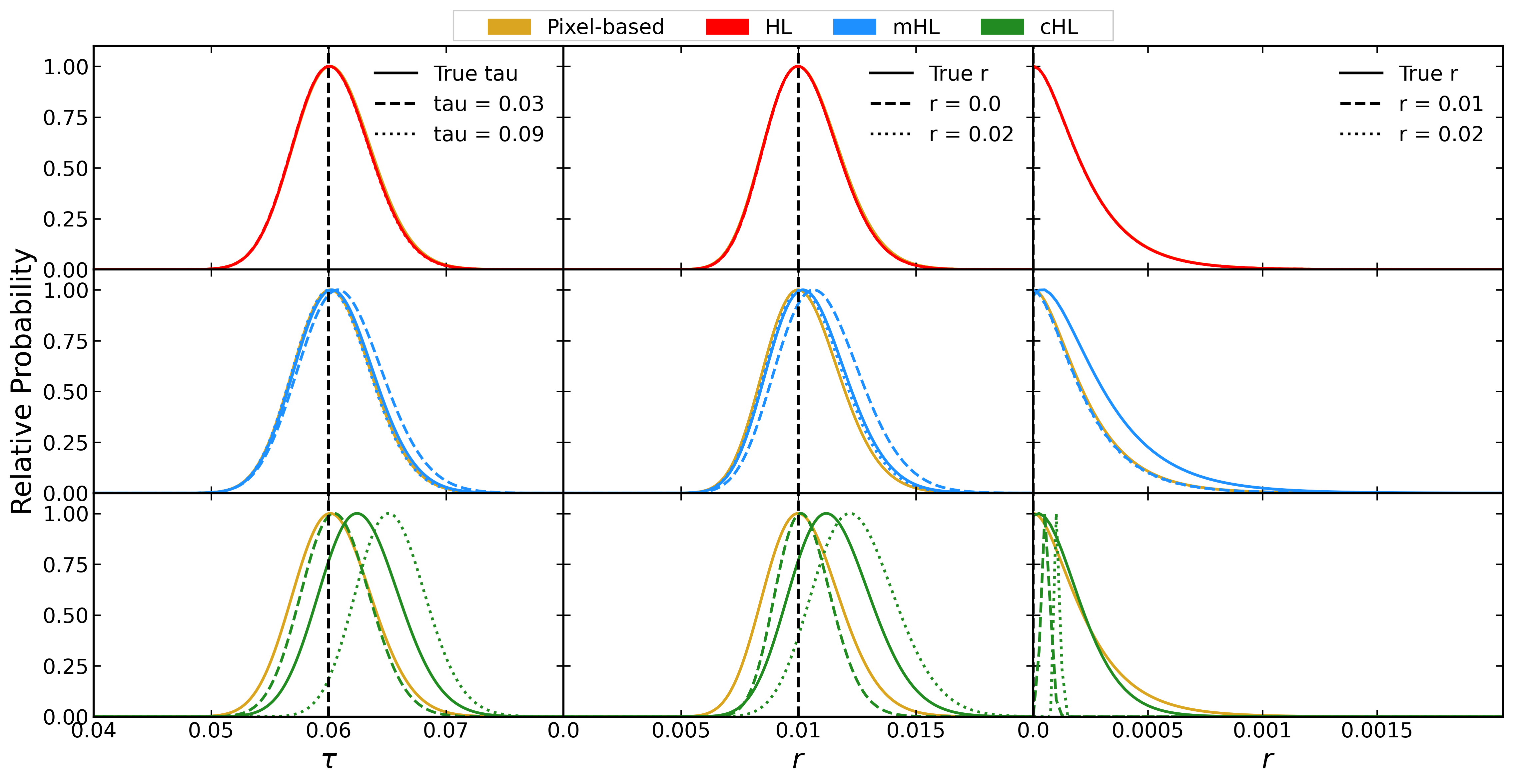} 
    \caption{Comparison between pixel-based (gold), HL (red), mHL (blue), and cHL (green) likelihoods while assuming different fiducial cosmological values in full-sky. The columns correspond to signal-dominated, tensor-detection, and noise-dominated cases. Dashed and dotted curves indicate a misestimation of the fiducial spectra. The specific values for the fiducial parameters are shown in each column. The black vertical lines refer to the true values of the parameters ($\tau = 0.06$, $\rpivot=0.01$ and $\rpivot=0$).} 
    \label{fig: wrongfid_BB_oHL_mHL}
\end{figure}

\autoref{fig: wrongfid_BB_oHL_mHL} presents the results for a misestimation of the fiducial spectrum. Note that here we focus on the likelihood effects of a wrong estimation, not on the power spectrum estimation step. 

The structure of the plot is analogous to \autoref{fig: mismatch_BB_oHL_mHL}, but here the dashed and dotted lines reflect errors in the fiducial assumptions for either $\tau$ or $\rpivot$.

Notably, the HL is extremely stable against such a mismatch. While this stability is expected \cite{hamimeche2008LikelihoodAnalysisCMBTemperature}, it is interesting to observe that it holds even for significantly erroneous fiducial values.

In contrast, the mHL demonstrates a discrete robustness against fiducial mismatches. In the noise-dominated case, when a wrong fiducial $\rpivot$, the posterior seems to collapse to the correct result, irrespective of the fiducial value assumed. However, this is just serendipitous. We tested other values of the fiducial between $\rpivot=0$ and $\rpivot=0.01$, and the posterior shifts continuously from the correct result to the dashed line in the plot.

On the other hand, the cHL likelihood shows a strong dependence on the assumed fiducial. This behavior is linked to the offset, which directly relies on the chosen fiducial spectrum. Indeed, focusing on the noise-dominated case, when $\rpivot$ is assumed to be $0.01$ or $0.02$, the offset gets significantly suppressed, to the point that on the very first multipoles it drops to zero. As a consequence, nothing is preventing negative values from generating numerical problems, that manifest here in two very sharp peaks.

It is worth reiterating that a mismatch in the fiducial spectrum can be ``detected'' and resolved iteratively. The key takeaway here is that the cHL requires careful mitigation of this effect, whereas the HL and mHL remain more stable in the presence of such mismatches.

\section{The role of the offset}\label{app: offset}
\begin{figure}
    \centering
    \includegraphics[width=\linewidth]{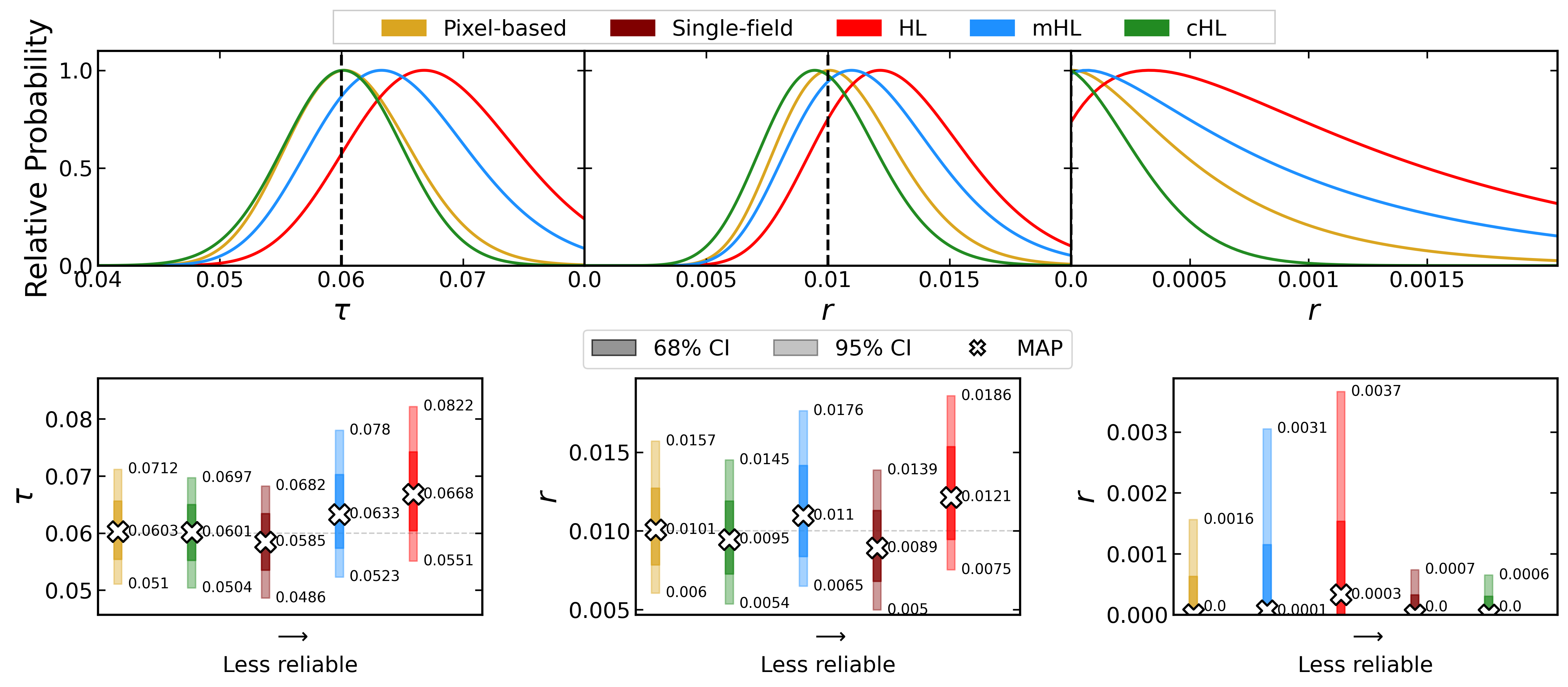}
    \caption{Comparison between pixel-based (gold), HL (red), mHL (blue), and cHL (green) likelihoods with $\fsky = 40\%$ omitting the offset. The columns correspond to signal-dominated, tensor-detection, and noise-dominated cases. The black vertical lines represent the true values of the parameters ($\tau = 0.06$, $\rpivot=0.01$ and $\rpivot=0$). The bottom row provides a quantitative comparison of the posteriors above, together with the single-field HL (maroon). Crosses show MAP estimates, while shaded bars indicate 68\% and 95\% credible intervals.}
    \label{fig: fsky40_mismatch_BB_oHL_mHL}
\end{figure}

Building on the results shown in \autoref{fig: offset_fsky40_mismatch_BB_oHL_mHL}, we now explore the role of the offset in the HL and mHL results. As mentioned in \autoref{sec: methodology}, the issue of negative spectra manifests in the HL transformation as negative eigenvalues in the matrix $\mathbf{C}_\ell^{-1/2}\hat{\mathbf{C}}_\ell \mathbf{C}_\ell^{-1/2}$. To mimic the oHL (and cHL), we add an offset to the diagonal elements of the various matrices involved to regularize their eigenvalues. This brings us to the main results of this work, i.e. \autoref{fig: offset_fsky40_mismatch_BB_oHL_mHL}, while \autoref{fig: fsky40_mismatch_BB_oHL_mHL} shows the equivalent results without adding an offset to the HL and the mHL \footnote{We emphasize that both the cHL and the single-field HL are always computed adding the corresponding offset.}.

Note that here we report solely the case in which $\widehat{N}_\ell = N_\ell^{\rm True}$ as the mismatch of the noise bias behaves as expected. Thus, we collapse the $3 \times 3$ grid of \autoref{fig: offset_fsky40_mismatch_BB_oHL_mHL} into a single row comparing directly the three approximations.

We observe that the cut-sky introduces a bias into the HL approximation, causing it to significantly diverge from the pixel-based likelihood. In fact, the partial sky acts as additional noise, shifting the distribution rightward.

The mHL likelihood also exhibits a bias rightward when applied to the cut-sky; however, this effect is less pronounced compared to the HL, as the ``extra noise'' induced by the partial sky is somewhat marginalized over. The only robust case is the noise-dominated case, where the peak of the posterior remains largely unchanged.

The cHL is already discussed in \autoref{sec: results}.

In the bottom line of the plot, we can see that the KL divergence ranks the distributions very differently from \autoref{fig: offset_fsky40_mismatch_BB_oHL_mHL}. In the signal-dominated case, the most reliable one is the cHL due to the strong bias experienced by the other two. In the tensor-detection case, the mHL and the cHL obtain a very similar KL divergence that slightly favors the mHL. Finally, in the noise-dominated case, although mHL and HL largely overestimate the uncertainty, they obtain a better KL divergence. Still, it is clear that none of the approximations is doing a great job in recovering the PB here.

\begin{figure}
    \centering
    \includegraphics[width=\linewidth]{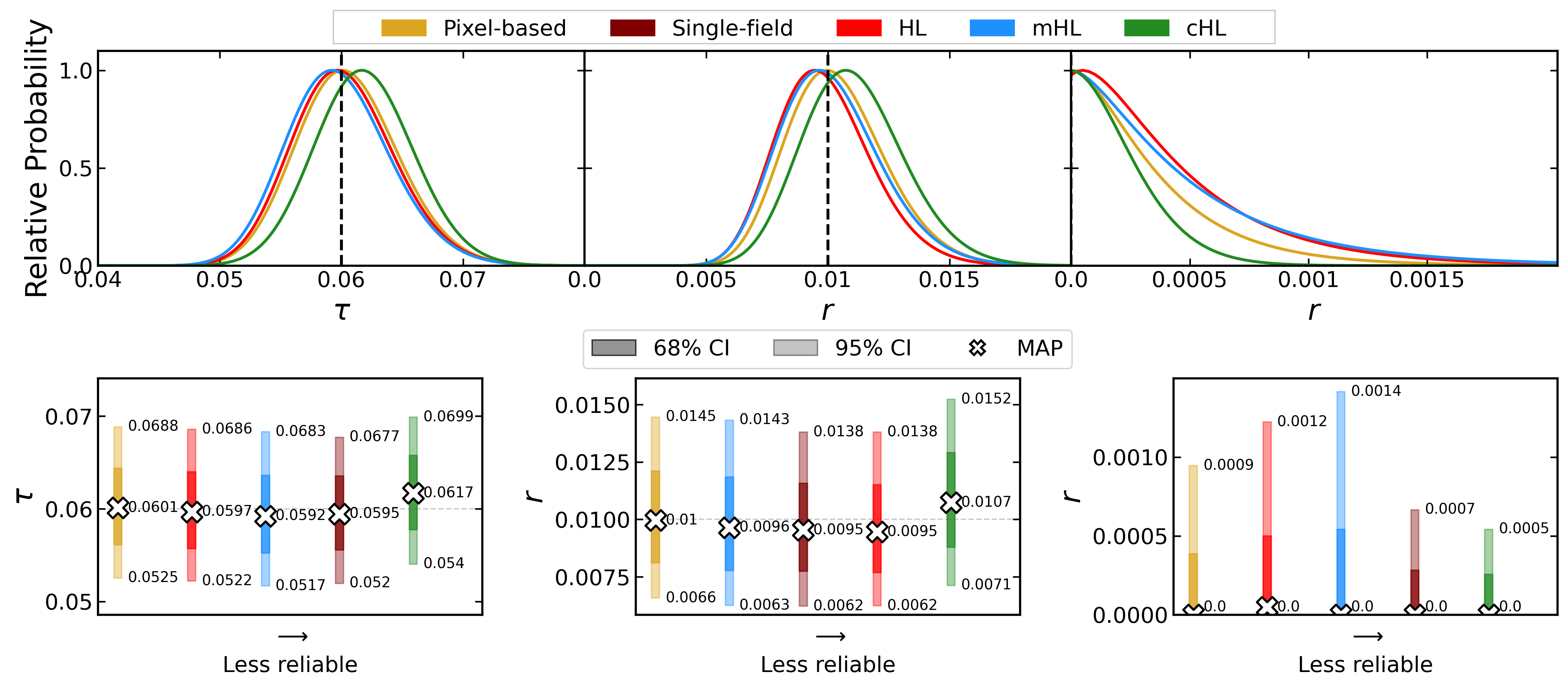}
    \caption{Comparison between pixel-based (gold), HL (red), mHL (blue), and cHL (green) likelihoods with $\fsky = 60\%$ omitting the offset. The columns correspond to signal-dominated, tensor-detection, and noise-dominated cases. The black vertical lines represent the true values of the parameters ($\tau = 0.06$, $\rpivot=0.01$ and $\rpivot=0$). The bottom row provides a quantitative comparison of the posteriors above, together with the single-field HL (maroon). Crosses show MAP estimates, while shaded bars indicate 68\% and 95\% credible intervals.}
    \label{fig: fsky60_mismatch_BB_oHL_mHL}
\end{figure}

As shown in \autoref{fig: fsky60_mismatch_BB_oHL_mHL}, for a less aggressive sky fraction of $\fsky = 60\%$, neither the HL nor the mHL shows very significant biases. This highlights a key strength of these likelihoods: for $\fsky \geq 60\%$, the power spectrum estimation is stable enough, and there is no need to define an offset to obtain unbiased results.

Here we can also notice that the cHL appears biased, while in the $40\%$ it was not. As mentioned before, the cHL is biased rightward in full-sky, but the mask is causing a pull leftward. For $\fsky = 40\%$, the two effects balance, while in this case with $\fsky = 60\%$, they do not.

Once again the KL divergence ranks differently the various approximations. In the signal-dominated case, the HL gives the best results without doubts. Whether one has to resort to using one of mHL and cHL, they give very similar scores and appear almost symmetric with respect to the true value of $\tau$. In the tensor-detection case, instead, the mHL gives the best agreement to the PB; finally, in the noise-dominated case, the most reliable is the HL, followed by the mHL, and both tend to overestimate the 95\% upper bound.

\subsection{Changing offset}

In \autoref{sec: methodology}, we anticipated that the offset used in this work follows the prescription of \cite{mangilli2015LargescaleCMBTemperaturepolarization}. While this reflects the original definition of the oHL, it is not the only possible way to compute this quantity. For instance, the \Planck~PR4 low-$\ell$ polarization results adopt a different approach within the so-called \texttt{LoLLiPoP} likelihood \cite{tristram2022ImprovedLimitsTensortoscalarratio}. Instead of requiring 99\% of spectra to be positive, this method compares the empirical variance of $C_\ell$ with the expected variance given by the Knox formula \cite{Knox:1995dq}

\begin{equation}
    \sigma^2_\ell = \frac{2 \qty(C_\ell + O_\ell)^2}{(2\ell+1)\fsky}\ ,
\end{equation}

and extracts the offset iteratively. We re-compute the posteriors using this alternative definition to test the sensitivity of our results to the offset choice. As an example, we consider the tensor-detection case.

\begin{figure}
    \centering
    \includegraphics[width=\linewidth]{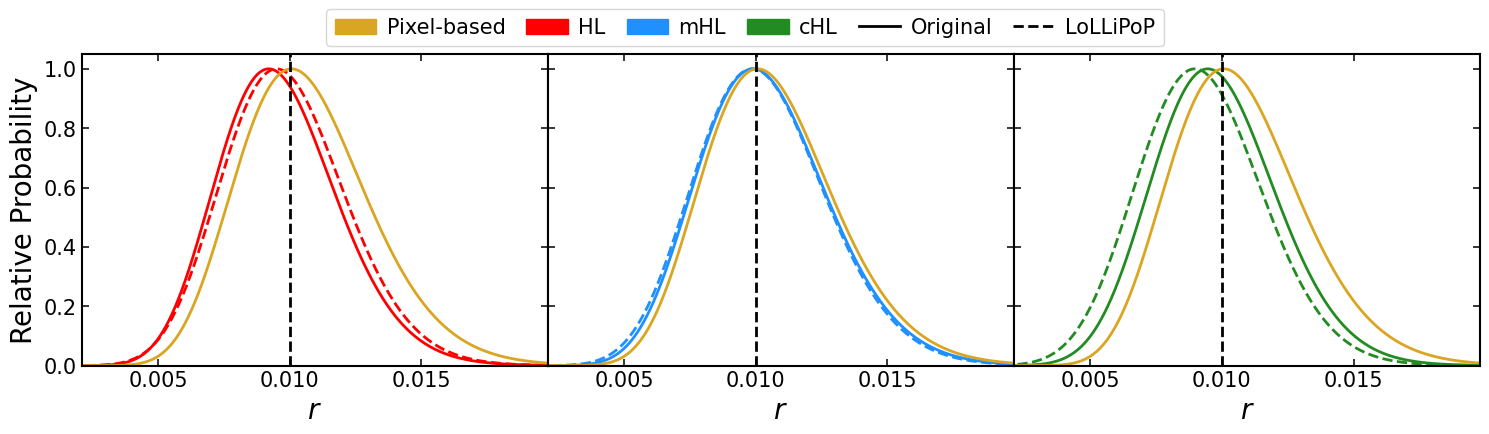}
    \caption{Comparison between pixel-based (gold), HL (red), mHL (blue), and cHL (green) likelihoods with $\fsky = 40\%$ obtained with different offsets. Each panel corresponds to signal-dominated, tensor-detection, and noise-dominated cases. Solid lines refer to the computation presented in \autoref{sec: methodology}, while the dashed lines correspond to the alternative definition introduced here. The black vertical lines indicate the true values of the parameters ($\tau = 0.06$, $\rpivot=0.01$ and $\rpivot=0$).
    }
    \label{fig: offsets}
\end{figure}

The posteriors for the different likelihoods are shown in \autoref{fig: offsets}. As discussed in \autoref{sec: results}, the mHL provides the best agreement with the PB. Since the offset is applied only to the auto-correlations, which are marginalized over in the mHL (see \autoref{sec: methodology}), its results are largely unaffected by the choice of offset. In contrast, both the HL and cHL show greater sensitivity to the offset computation.

Overall, a larger offset appears to ``Gaussianize'' the distributions, making them less skewed. Although we do not show it explicitly, the LoLLiPoP-based prescription consistently produces larger offsets than the original method from \cite{mangilli2015LargescaleCMBTemperaturepolarization} in the tensor-detection case.

A more detailed study of how different offset definitions affect the analysis is left for future work.

\section{Approximating the likelihood with a Gaussian}\label{app: gaussian}

A common approximation for the distribution of $C_\ell$ values is the Gaussian. In this case, the likelihood has the same form as \autoref{eq:hl_likelihood}, where this time the $X_\ell$ vector is the concatenation of the $n(n+1)/2$ independent $(\hat{C}_\ell - C_\ell)$ components. This approach is justified at high multipoles, where the number of degrees of freedom is large enough for the central limit theorem to apply. However, it is sometimes also used at large scales, where the goal is to approximate the effect of the mask through the covariance.

\begin{figure}
    \centering
    \includegraphics[width=\linewidth]{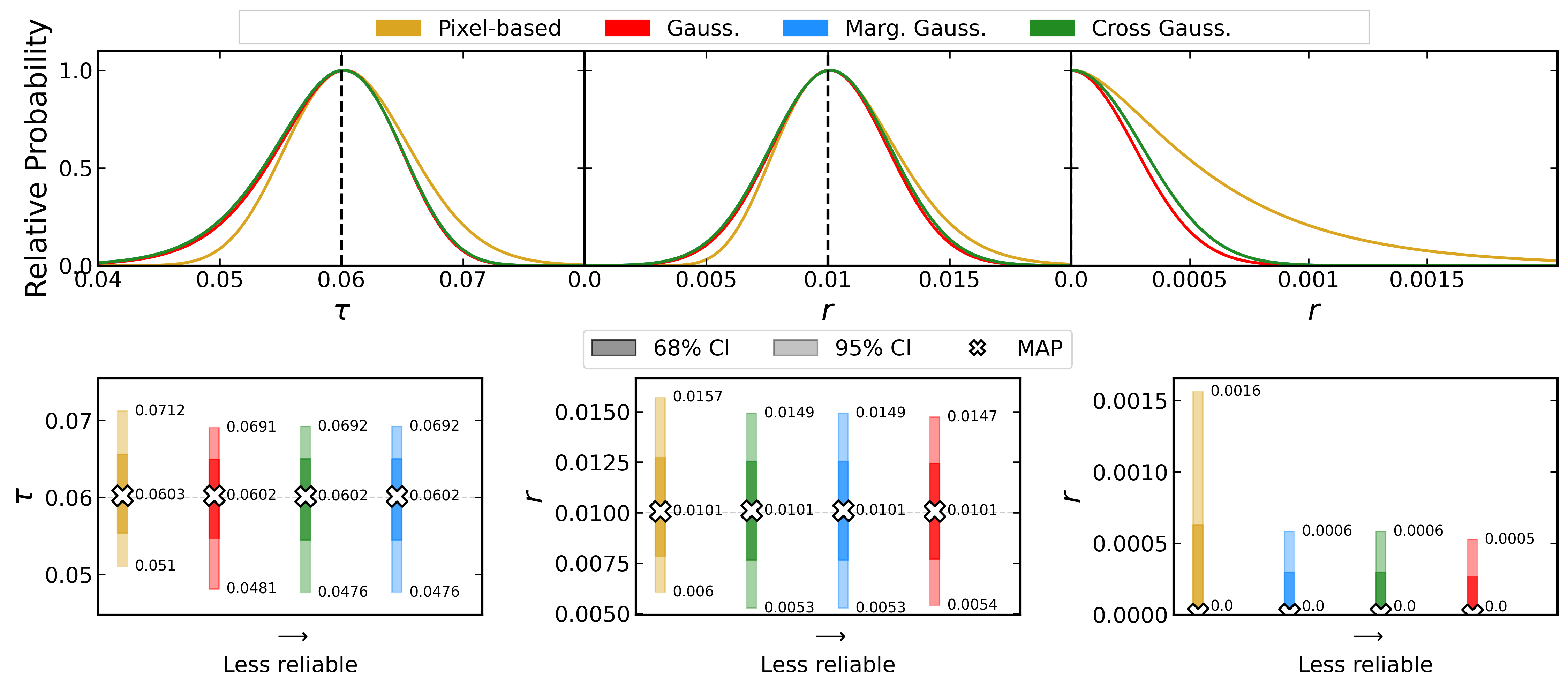}
    \caption{Comparison between pixel-based (gold), Gaussian (red), marginalized Gaussian (blue), and cross-spectra Gaussian (green) likelihoods with $\fsky = 40\%$. The columns correspond to signal-dominated, tensor-detection, and noise-dominated cases. The black vertical lines represent the true values of the parameters ($\tau = 0.06$, $\rpivot=0.01$, and $\rpivot=0$). The bottom row provides a quantitative comparison of the posteriors above: crosses show MAP estimates, while shaded bars indicate 68\% and 95\% credible intervals.}
    \label{fig: gaussian_case}
\end{figure}

\autoref{fig: gaussian_case} shows the results. The naming convention here matches the HL-based case; however, the marginalized Gaussian and cross-spectra Gaussian approximations are now identical: without the HL transformation, the two are mathematically the same, and this is confirmed by the results. The difference with the Gaussian lies in the inclusion of auto-spectra.

All posteriors are unbiased in this case. Since the data vector is the direct difference between the observed and model spectra, it is harder to get biased results when using an unbiased estimator like the QML.

However, the Gaussian approximation fails to capture the skewed shape of the true distribution. This leads to underestimated uncertainties in the noise-dominated case, and distorted distributions for $\tau$ and $\rpivot$ in the signal-dominated and tensor-detection cases, favoring low values.

These results emphasize that non-Gaussian approximations, such as those in the ``HL family'', remain necessary for accurate parameter inference on large scales.

\end{appendix}

\printbibliography

@article{aghanimPlanck2018Results2020a,
  ids = {Planck_like},
  title = {Planck 2018 Results. {{V}}. {{CMB}} Power Spectra and Likelihoods},
  author = {Aghanim, N. and others},
  year = {2020},
  journal = {Astronomy and Astrophysics},
  volume = {641},
  eprint = {1907.12875},
  primaryclass = {astro-ph.CO},
  pages = {A5},
  doi = {10.1051/0004-6361/201936386},
  archiveprefix = {arxiv},
  collaboration = {Planck}
}

@article{boggessCOBEMissionIts1992,
  ids = {COBE_mission},
  title = {The {{COBE}} Mission - {{Its}} Design and Performance Two Years after Launch},
  author = {Boggess, N.W. and others},
  year = {1992},
  journal = {Astrophysical Journal},
  volume = {397},
  pages = {420--429},
  doi = {10.1086/171797}
}

@article{caronesAnalysisNILCPerformance2022,
  author = "Carones, Alessandro and Migliaccio, Marina and Marinucci, Domenico and Vittorio, Nicola",
  title = "{Analysis of Needlet Internal Linear Combination performance on B-mode data from sub-orbital experiments}",
  eprint = "2208.12059",
  archivePrefix = "arXiv",
  primaryClass = "astro-ph.CO",
  doi = "10.1051/0004-6361/202244824",
  journal = "Astron. Astrophys.",
  volume = "677",
  pages = "A147",
  year = "2023"
}

@article{collaboration2021BICEPKeckXIIIImproved,
  ids = {BICEP_2021},
  title = {{{BICEP}} / {{Keck XIII}}: {{Improved Constraints}} on {{Primordial Gravitational Waves}} Using {{Planck}}, {{WMAP}}, and {{BICEP}}/{{Keck Observations}} through the 2018 {{Observing Season}}},
  shorttitle = {{{BICEP}} / {{Keck XIII}}},
  author = {Ade, P. A. R. and Ahmed, Z. and Amiri, M. and Barkats, D. and Thakur, R. Basu and Beck, D. and Bischoff, C. and Bock, J. J. and Boenish, H. and Bullock, E. and Buza, V. and Cheshire IV, J. R. and Connors, J. and Cornelison, J. and Crumrine, M. and Cukierman, A. and Denison, E. V. and Dierickx, M. and Duband, L. and Eiben, M. and Fatigoni, S. and Filippini, J. P. and Fliescher, S. and {Goeckner-Wald}, N. and Goldfinger, D. C. and Grayson, J. and Grimes, P. and Halal, G. and Hall, G. and Halpern, M. and Hand, E. and Harrison, S. and Henderson, S. and Hildebrandt, S. R. and Hilton, G. C. and Hubmayr, J. and Hui, H. and Irwin, K. D. and Kang, J. and Karkare, K. S. and Karpel, E. and Kefeli, S. and Kernasovskiy, S. A. and Kovac, J. M. and Kuo, C. L. and Lau, K. and Leitch, E. M. and Lennox, A. and Megerian, K. G. and Minutolo, L. and Moncelsi, L. and Nakato, Y. and Namikawa, T. and Nguyen, H. T. and O'Brient, R. and Ogburn IV, R. W. and Palladino, S. and Prouve, T. and Pryke, C. and Racine, B. and Reintsema, C. D. and Richter, S. and Schillaci, A. and Schmitt, B. L. and Schwarz, R. and Sheehy, C. D. and Soliman, A. and Germaine, T. St and Steinbach, B. and Sudiwala, R. V. and Teply, G. P. and Thompson, K. L. and Tolan, J. E. and Tucker, C. and Turner, A. and Umilta, C. and Verges, C. and Vieregg, A. G. and Wandui, A. and Weber, A. C. and Wiebe, D. V. and Willmert, J. and Wong, C. L. and Wu, W. L. K. and Yang, H. and Yoon, K. W. and Young, E. and Yu, C. and Zeng, L. and Zhang, C. and Zhang, S.},
  year = {2021},
  month = oct,
  journal = {Physical Review Letters},
  volume = {127},
  number = {15},
  eprint = {2110.00483},
  primaryclass = {astro-ph},
  pages = {151301},
  issn = {0031-9007, 1079-7114},
  doi = {10.1103/PhysRevLett.127.151301},
  visitdate = {2022-11-24},
  archiveprefix = {arxiv},
  collaboration = {BICEP/Keck}
}

@article{gerbinoLikelihoodMethodsCMB2020,
  ids = {gerbino_likelihood_2020},
  title = {Likelihood Methods for {{CMB}} Experiments},
  author = {Gerbino, Martina and Lattanzi, Massimiliano and Migliaccio, Marina and Pagano, Luca and Salvati, Laura and Colombo, Loris and Gruppuso, Alessandro and Natoli, Paolo and Polenta, Gianluca},
  year = {2020},
  month = feb,
  journal = {Frontiers in Physics},
  volume = {8},
  eprint = {1909.09375},
  primaryclass = {astro-ph},
  pages = {15},
  issn = {2296-424X},
  doi = {10.3389/fphy.2020.00015},
  visitdate = {2022-11-24},
  archiveprefix = {arxiv}
}

@article{gorskiHEALPixFrameworkHighResolution2005,
  ids = {healpix},
  title = {{{HEALPix}}: {{A Framework}} for {{High-Resolution Discretization}} and {{Fast Analysis}} of {{Data Distributed}} on the {{Sphere}}},
  author = {G{\'o}rski, K. M. and Hivon, E. and Banday, A. J. and {Wand elt}, B. D. and Hansen, F. K. and Reinecke, M. and Bartelmann, M.},
  year = {2005},
  month = apr,
  journal = {Astrophysical Journal},
  volume = {622},
  number = {2},
  eprint = {astro-ph/0409513},
  pages = {759--771},
  doi = {10.1086/427976},
  archiveprefix = {arxiv}
}

@article{hamimeche2008LikelihoodAnalysisCMBTemperature,
  ids = {Hamimeche_2008},
  title = {Likelihood {{Analysis}} of {{CMB Temperature}} and {{Polarization Power Spectra}}},
  author = {Hamimeche, Samira and Lewis, Antony},
  year = {2008},
  month = may,
  journal = {Physical Review D},
  volume = {77},
  number = {10},
  eprint = {0801.0554},
  primaryclass = {astro-ph},
  pages = {103013},
  issn = {1550-7998, 1550-2368},
  doi = {10.1103/PhysRevD.77.103013},
  visitdate = {2022-11-24},
  archiveprefix = {arxiv}
}

@article{harrisArrayProgrammingNumPy2020,
  ids = {numpy},
  title = {Array Programming with {{NumPy}}},
  author = {Harris, Charles R. and Millman, K. Jarrod and {van der Walt}, St{\'e}fan J and Gommers, Ralf and Virtanen, Pauli and Cournapeau, David and Wieser, Eric and Taylor, Julian and Berg, Sebastian and Smith, Nathaniel J. and Kern, Robert and Picus, Matti and Hoyer, Stephan and {van Kerkwijk}, Marten H. and Brett, Matthew and Haldane, Allan and {Fern{\'a}ndez del R{\'i}o}, Jaime and Wiebe, Mark and Peterson, Pearu and {G{\'e}rard-Marchant}, Pierre and Sheppard, Kevin and Reddy, Tyler and Weckesser, Warren and Abbasi, Hameer and Gohlke, Christoph and Oliphant, Travis E.},
  year = {2020},
  journal = {Nature},
  volume = {585},
  eprint = {2006.10256},
  pages = {357--362},
  doi = {10.1038/s41586-020-2649-2}
}

@article{hunterMatplotlib2DGraphics2007,
  ids = {matplotlib},
  title = {Matplotlib: {{A 2D}} Graphics Environment},
  author = {Hunter, J. D.},
  year = {2007},
  journal = {Computing in Science \& Engineering},
  volume = {9},
  number = {3},
  pages = {90--95},
  publisher = {{IEEE COMPUTER SOC}},
  doi = {10.1109/MCSE.2007.55}
}

@article{huWanderingBackgroundCMB1995,
  ids = {Hu:1995em},
  title = {Wandering in the {{Background}}: {{A CMB Explorer}}},
  author = {Hu, Wayne},
  year = {1995},
  month = aug,
  journal = {arXiv e-prints},
  eprint = {astro-ph/9508126},
  pages = {astro-ph/9508126},
  archiveprefix = {arxiv}
}

@article{kamionkowskiQuestModesInflationary2016,
  ids = {Kamionkowski_2016},
  title = {The {{Quest}} for {{B Modes}} from {{Inflationary Gravitational Waves}}},
  author = {Kamionkowski, Marc and Kovetz, Ely D.},
  year = {2016},
  month = sep,
  journal = {Annual Review of Astronomy and Astrophysics},
  volume = {54},
  number = {1},
  eprint = {1510.06042},
  primaryclass = {astro-ph, physics:gr-qc, physics:hep-ph, physics:hep-th},
  pages = {227--269},
  issn = {0066-4146, 1545-4282},
  doi = {10.1146/annurev-astro-081915-023433},
  visitdate = {2023-06-19},
  archiveprefix = {arxiv}
}

@article{kamionkowskiStatisticsCosmicMicrowave1997,
  ids = {Kamionkowski_1997},
  title = {Statistics of Cosmic Microwave Background Polarization},
  author = {Kamionkowski, Marc and Kosowsky, Arthur and Stebbins, Albert},
  year = {1997},
  month = jun,
  journal = {Physical Review D},
  volume = {55},
  number = {12},
  eprint = {astro-ph/9611125},
  pages = {7368--7388},
  publisher = {{American Physical Society (APS)}},
  doi = {10.1103/physrevd.55.7368}
}

@inproceedings{lam2015NumbaLLVMbasedPythonJIT,
  author = {Lam, Siu Kwan and Pitrou, Antoine and Seibert, Stanley},
  title = {Numba: a LLVM-based Python JIT compiler},
  year = {2015},
  isbn = {9781450340052},
  publisher = {Association for Computing Machinery},
  address = {New York, NY, USA},
  doi = {10.1145/2833157.2833162},
  booktitle = {Proceedings of the Second Workshop on the LLVM Compiler Infrastructure in HPC},
  articleno = {7},
  numpages = {6},
  location = {Austin, Texas},
  series = {LLVM '15}
}

@article{lewisEfficientComputationCosmic2000,
  ids = {Lewis_2000},
  title = {Efficient {{Computation}} of {{Cosmic Microwave Background Anisotropies}} in {{Closed Friedmann-Robertson-Walker Models}}},
  author = {Lewis, Antony and Challinor, Anthony and Lasenby, Anthony},
  year = {2000},
  month = aug,
  journal = {The Astrophysical Journal},
  volume = {538},
  number = {2},
  eprint = {astro-ph/9911177},
  pages = {473--476},
  publisher = {{American Astronomical Society}},
  doi = {10.1086/309179}
}

@article{mangilli2015LargescaleCMBTemperaturepolarization,
  ids = {Mangilli:2015xya},
  title = {Large-Scale {{CMB}} Temperature and Polarization Cross-Spectra Likelihoods},
  author = {Mangilli, A. and Plaszczynski, S. and Tristram, M.},
  year = {2015},
  month = nov,
  journal = {Monthly Notices of the Royal Astronomical Society},
  volume = {453},
  number = {3},
  eprint = {1503.01347},
  primaryclass = {astro-ph},
  pages = {3175--3190},
  issn = {0035-8711, 1365-2966},
  doi = {10.1093/mnras/stv1733},
  visitdate = {2022-11-24},
  archiveprefix = {arxiv}
}

@article{planckcollaboration2020Planck2018ResultsVI,
  ids = {Planck_parameters},
  title = {Planck 2018 Results. {{VI}}. {{Cosmological}} Parameters},
  author = {Aghanim, N. and Akrami, Y. and Ashdown, M. and Aumont, J. and Baccigalupi, C. and Ballardini, M. and Banday, A. J. and Barreiro, R. B. and Bartolo, N. and Basak, S. and Battye, R. and Benabed, K. and Bernard, J.-P. and Bersanelli, M. and Bielewicz, P. and Bock, J. J. and Bond, J. R. and Borrill, J. and Bouchet, F. R. and Boulanger, F. and Bucher, M. and Burigana, C. and Butler, R. C. and Calabrese, E. and Cardoso, J.-F. and Carron, J. and Challinor, A. and Chiang, H. C. and Chluba, J. and Colombo, L. P. L. and Combet, C. and Contreras, D. and Crill, B. P. and Cuttaia, F. and {de Bernardis}, P. and {de Zotti}, G. and Delabrouille, J. and Delouis, J.-M. and Di Valentino, E. and Diego, J. M. and Dor{\'e}, O. and Douspis, M. and Ducout, A. and Dupac, X. and Dusini, S. and Efstathiou, G. and Elsner, F. and En{\ss}lin, T. A. and Eriksen, H. K. and Fantaye, Y. and Farhang, M. and Fergusson, J. and {Fernandez-Cobos}, R. and Finelli, F. and Forastieri, F. and Frailis, M. and Fraisse, A. A. and Franceschi, E. and Frolov, A. and Galeotta, S. and Galli, S. and Ganga, K. and {G{\'e}nova-Santos}, R. T. and Gerbino, M. and Ghosh, T. and {Gonz{\'a}lez-Nuevo}, J. and G{\'o}rski, K. M. and Gratton, S. and Gruppuso, A. and Gudmundsson, J. E. and Hamann, J. and Handley, W. and Hansen, F. K. and Herranz, D. and Hildebrandt, S. R. and Hivon, E. and Huang, Z. and Jaffe, A. H. and Jones, W. C. and Karakci, A. and Keih{\"a}nen, E. and Keskitalo, R. and Kiiveri, K. and Kim, J. and Kisner, T. S. and Knox, L. and Krachmalnicoff, N. and Kunz, M. and {Kurki-Suonio}, H. and Lagache, G. and Lamarre, J.-M. and Lasenby, A. and Lattanzi, M. and Lawrence, C. R. and Jeune, M. Le and Lemos, P. and Lesgourgues, J. and Levrier, F. and Lewis, A. and Liguori, M. and Lilje, P. B. and Lilley, M. and Lindholm, V. and {L{\'o}pez-Caniego}, M. and Lubin, P. M. and Ma, Y.-Z. and {Mac{\'i}as-P{\'e}rez}, J. F. and Maggio, G. and Maino, D. and Mandolesi, N. and Mangilli, A. and {Marcos-Caballero}, A. and Maris, M. and Martin, P. G. and Martinelli, M. and {Mart{\'i}nez-Gonz{\'a}lez}, E. and Matarrese, S. and Mauri, N. and McEwen, J. D. and Meinhold, P. R. and Melchiorri, A. and Mennella, A. and Migliaccio, M. and Millea, M. and Mitra, S. and {Miville-Desch{\^e}nes}, M.-A. and Molinari, D. and Montier, L. and Morgante, G. and Moss, A. and Natoli, P. and {N{\o}rgaard-Nielsen}, H. U. and Pagano, L. and Paoletti, D. and Partridge, B. and Patanchon, G. and Peiris, H. V. and Perrotta, F. and Pettorino, V. and Piacentini, F. and Polastri, L. and Polenta, G. and Puget, J.-L. and Rachen, J. P. and Reinecke, M. and Remazeilles, M. and Renzi, A. and Rocha, G. and Rosset, C. and Roudier, G. and {Rubi{\~n}o-Mart{\'i}n}, J. A. and {Ruiz-Granados}, B. and Salvati, L. and Sandri, M. and Savelainen, M. and Scott, D. and Shellard, E. P. S. and Sirignano, C. and Sirri, G. and Spencer, L. D. and Sunyaev, R. and {Suur-Uski}, A.-S. and Tauber, J. A. and Tavagnacco, D. and Tenti, M. and Toffolatti, L. and Tomasi, M. and Trombetti, T. and Valenziano, L. and Valiviita, J. and Van Tent, B. and Vibert, L. and Vielva, P. and Villa, F. and Vittorio, N. and Wandelt, B. D. and Wehus, I. K. and White, M. and White, S. D. M. and Zacchei, A. and Zonca, A.},
  year = {2020},
  month = sep,
  journal = {Astronomy \& Astrophysics},
  volume = {641},
  eprint = {1807.06209},
  primaryclass = {astro-ph},
  pages = {A6},
  issn = {0004-6361, 1432-0746},
  doi = {10.1051/0004-6361/201833910},
  visitdate = {2022-11-24},
  archiveprefix = {arxiv},
  collaboration = {Planck}
}

@article{PTEP_LiteBIRD,
  title = {Probing Cosmic Inflation with the {{LiteBIRD}} Cosmic Microwave Background Polarization Survey},
  author = {Allys, E. and Arnold, K. and Aumont, J. and Aurlien, R. and Azzoni, S. and Baccigalupi, C. and Banday, A. J. and Banerji, R. and Barreiro, R. B. and Bartolo, N. and Bautista, L. and Beck, D. and Beckman, S. and Bersanelli, M. and Boulanger, F. and Brilenkov, M. and Bucher, M. and Calabrese, E. and Campeti, P. and Carones, A. and Casas, F. J. and Catalano, A. and Chan, V. and Cheung, K. and Chinone, Y. and Clark, S. E. and Columbro, F. and D'Alessandro, G. and {de Bernardis}, P. and {de Haan}, T. and {de la Hoz}, E. and De Petris, M. and Della Torre, S. and {Diego-Palazuelos}, P. and Dotani, T. and Duval, J. M. and Elleflot, T. and Eriksen, H. K. and Errard, J. and {Essinger-Hileman}, T. and Finelli, F. and Flauger, R. and Franceschet, C. and Fuskeland, U. and Galloway, M. and Ganga, K. and Gerbino, M. and Gervasi, M. and {G{\'e}nova-Santos}, R. T. and Ghigna, T. and Giardiello, S. and Gjerl{\o}w, E. and Grain, J. and Grupp, F. and Gruppuso, A. and Gudmundsson, J. E. and Halverson, N. W. and Hargrave, P. and Hasebe, T. and Hasegawa, M. and Hazumi, M. and {Henrot-Versill{\'e}}, S. and Hensley, B. and Hergt, L. T. and Herman, D. and Hivon, E. and Hlozek, R. A. and Hornsby, A. L. and Hoshino, Y. and Hubmayr, J. and Ichiki, K. and Iida, T. and Imada, H. and Ishino, H. and Jaehnig, G. and Katayama, N. and Kato, A. and Keskitalo, R. and Kisner, T. and Kobayashi, Y. and Kogut, A. and Kohri, K. and Komatsu, E. and Komatsu, K. and Konishi, K. and Krachmalnicoff, N. and Kuo, C. L. and Lamagna, L. and Lattanzi, M. and Lee, A. T. and Leloup, C. and Levrier, F. and Linder, E. and Luzzi, G. and {Macias-Perez}, J. and Maffei, B. and Maino, D. and Mandelli, S. and {Mart{\'i}nez-Gonz{\'a}lez}, E. and Masi, S. and Massa, M. and Matarrese, S. and Matsuda, F. T. and Matsumura, T. and Mele, L. and Migliaccio, M. and Minami, Y. and Moggi, A. and Montgomery, J. and Montier, L. and Morgante, G. and Mot, B. and Nagano, Y. and Nagasaki, T. and Nagata, R. and Nakano, R. and Namikawa, T. and Nati, F. and Natoli, P. and Nerval, S. and Noviello, F. and Odagiri, K. and Oguri, S. and Ohsaki, H. and Pagano, L. and Paiella, A. and Paoletti, D. and Passerini, A. and Patanchon, G. and Piacentini, F. and Piat, M. and Polenta, G. and Poletti, D. and Prouv{\'e}, T. and Puglisi, G. and Rambaud, D. and Raum, C. and Realini, S. and Reinecke, M. and Remazeilles, M. and Ritacco, A. and Roudil, G. and {Rubino-Martin}, J. A. and Russell, M. and Sakurai, H. and Sakurai, Y. and Sasaki, M. and Scott, D. and Sekimoto, Y. and Shinozaki, K. and Shiraishi, M. and Shirron, P. and Signorelli, G. and Spinella, F. and Stever, S. and Stompor, R. and Sugiyama, S. and Sullivan, R. M. and Suzuki, A. and Svalheim, T. L. and Switzer, E. and Takaku, R. and Takakura, H. and Takase, Y. and Tartari, A. and Terao, Y. and Thermeau, J. and Thommesen, H. and Thompson, K. L. and Tomasi, M. and Tominaga, M. and Tristram, M. and Tsuji, M. and Tsujimoto, M. and Vacher, L. and Vielva, P. and Vittorio, N. and Wang, W. and Watanuki, K. and Wehus, I. K. and Weller, J. and Westbrook, B. and Wilms, J. and Wollack, E. J. and Yumoto, J. and Zannoni, M.},
  year = {2022},
  eprint = {2202.02773},
  publisher = {{arXiv}},
  doi = {10.48550/ARXIV.2202.02773},
  collaboration = {LiteBIRD},
  copyright = {arXiv.org perpetual, non-exclusive license}
}

@article{seljakLineofSightIntegrationApproach1996,
  ids = {Seljak_1996},
  title = {A {{Line-of-Sight Integration Approach}} to {{Cosmic Microwave Background Anisotropies}}},
  author = {Seljak, Uros and Zaldarriaga, Matias},
  year = {1996},
  month = oct,
  journal = {The Astrophysical Journal},
  volume = {469},
  eprint = {astro-ph/9603033},
  pages = {437},
  publisher = {{American Astronomical Society}},
  issn = {1538-4357},
  doi = {10.1086/177793}
}

@article{seljakSignatureGravityWaves1997,
  ids = {Seljak_1997},
  title = {Signature of {{Gravity Waves}} in the {{Polarization}} of the {{Microwave Background}}},
  author = {Seljak, Uros and Zaldarriaga, Matias},
  year = {1997},
  month = mar,
  journal = {Physical Review Letters},
  volume = {78},
  number = {11},
  eprint = {astro-ph/9609169},
  pages = {2054--2057},
  publisher = {{American Physical Society (APS)}},
  doi = {10.1103/physrevlett.78.2054}
}

@article{smootCOBEObservationsResults1999,
  ids = {Smoot_1999},
  title = {{{COBE}} Observations and Results},
  author = {Smoot, George F.},
  year = {1999},
  journal = {Conference on 3K cosmology},
  publisher = {{ASCE}},
  doi = {10.1063/1.59326},
  isbn = {1563968479}
}

@article{tegmark1997HowMeasureCMBpower,
  title = {How to Measure {{CMB}} Power Spectra without Losing Information},
  author = {Tegmark, Max},
  year = {1997},
  month = may,
  journal = {Physical Review D},
  volume = {55},
  number = {10},
  eprint = {astro-ph/9611174},
  pages = {5895--5907},
  issn = {0556-2821, 1089-4918},
  doi = {10.1103/PhysRevD.55.5895},
  visitdate = {2023-05-29},
  archiveprefix = {arxiv}
}

@article{tristram2022ImprovedLimitsTensortoscalarratio,
  ids = {Tristram:2022},
  title = {Improved Limits on the Tensor-to-Scalar Ratio Using {{BICEP}} and {{Planck}} Data},
  author = {Tristram, M. and others},
  year = {2022},
  journal = {Physical Review D: Particles and Fields},
  volume = {105},
  number = {8},
  eprint = {2112.07961},
  primaryclass = {astro-ph.CO},
  pages = {083524},
  doi = {10.1103/PhysRevD.105.083524},
  archiveprefix = {arxiv}
}

@article{Tristram:2023haj,
  ids = {tristram2023CosmologicalParametersDerivedFinal,tristramCosmologicalParametersDerived2023},
  title = {Cosmological Parameters Derived from the Final ({{PR4}}) {{Planck}} Data Release},
  author = {Tristram, M. and Banday, A. J. and Douspis, M. and Garrido, X. and G{\'o}rski, K. M. and {Henrot-Versill{\'e}}, S. and Hergt, L. T. and Ili{\'c}, S. and Keskitalo, R. and Lagache, G. and Lawrence, C. R. and Partridge, B. and Scott, D.},
  year = {2024},
  journal = {Astronomy \& Astrophysics},
  eprint = {2309.10034},
  primaryclass = {astro-ph.CO},
  issn = {0004-6361, 1432-0746},
  doi = {10.1051/0004-6361/202348015},
  archiveprefix = {arxiv}
}

@article{virtanen2020SciPyFundamentalAlgorithmsscientific,
  title = {{{SciPy}} 1.0: {{Fundamental}} Algorithms for Scientific Computing in {{Python}}},
  shorttitle = {{{SciPy}} 1.0},
  author = {Virtanen, Pauli and Gommers, Ralf and Oliphant, Travis E. and Haberland, Matt and Reddy, Tyler and Cournapeau, David and Burovski, Evgeni and Peterson, Pearu and Weckesser, Warren and Bright, Jonathan and {van der Walt}, St{\'e}fan J. and Brett, Matthew and Wilson, Joshua and Millman, K. Jarrod and Mayorov, Nikolay and Nelson, Andrew R. J. and Jones, Eric and Kern, Robert and Larson, Eric and Carey, C. J. and Polat, {\.I}lhan and Feng, Yu and Moore, Eric W. and VanderPlas, Jake and Laxalde, Denis and Perktold, Josef and Cimrman, Robert and Henriksen, Ian and Quintero, E. A. and Harris, Charles R. and Archibald, Anne M. and Ribeiro, Ant{\^o}nio H. and Pedregosa, Fabian and {van Mulbregt}, Paul},
  year = {2020},
  month = mar,
  journal = {Nature Methods},
  volume = {17},
  number = {3},
  eprint = {1907.10121},
  pages = {261--272},
  publisher = {{Nature Publishing Group}},
  issn = {1548-7105},
  doi = {10.1038/s41592-019-0686-2},
  visitdate = {2022-12-28},
  copyright = {2020 The Author(s)},
  langid = {english}
}

@article{eriksen2008JointBayesianComponent,
  title = {Joint {{Bayesian Component Separation}} and {{CMB Power Spectrum Estimation}}},
  author = {Eriksen, H. K. and Jewell, J. B. and Dickinson, C. and Banday, A. J. and G{\'o}rski, K. M. and Lawrence, C. R.},
  year = {2008},
  journal = {The Astrophysical Journal},
  volume = {676},
  eprint = {0709.1058},
  primaryclass = {astro-ph},
  pages = {10--32},
  issn = {0004-637X},
  doi = {10.1086/525277},
  archiveprefix = {arxiv}
}

@article{stolyarov2002AllskyComponentSeparation,
  title = {All-Sky Component Separation for the {{Planck}} Mission},
  author = {Stolyarov, V. and Hobson, M. P. and Ashdown, M. A. J. and Lasenby, A. N.},
  year = {2002},
  journal = {Monthly Notices of the Royal Astronomical Society},
  volume = {336},
  number = {1},
  eprint = {astro-ph/0105432},
  pages = {97--111},
  issn = {0035-8711},
  doi = {10.1046/j.1365-8711.2002.05683.x},
  archiveprefix = {arxiv}
}

@article{stompor2009MaximumLikelihoodAlgorithm,
  title = {Maximum Likelihood Algorithm for Parametric Component Separation in Cosmic Microwave Background Experiments},
  author = {Stompor, Radek and Leach, Samuel and Stivoli, Federico and Baccigalupi, Carlo},
  year = {2009},
  journal = {Monthly Notices of the Royal Astronomical Society},
  volume = {392},
  eprint = {0804.2645},
  primaryclass = {astro-ph},
  pages = {216},
  issn = {0035-8711},
  doi = {10.1111/j.1365-2966.2008.14023.x},
  archiveprefix = {arxiv}
}

@article{bennett2003FirstYearWilkinsonMicrowave,
  title = {First-{{Year Wilkinson Microwave Anisotropy Probe}} ({{WMAP}}) {{Observations}}: {{Foreground Emission}}},
  shorttitle = {First-{{Year Wilkinson Microwave Anisotropy Probe}} ({{WMAP}}) {{Observations}}},
  author = {Bennett, C. L. and Hill, R. S. and Hinshaw, G. and Nolta, M. R. and Odegard, N. and Page, L. and Spergel, D. N. and Weiland, J. L. and Wright, E. L. and Halpern, M. and Jarosik, N. and Kogut, A. and Limon, M. and Meyer, S. S. and Tucker, G. S. and Wollack, E.},
  year = {2003},
  journal = {The Astrophysical Journal Supplement Series},
  volume = {148},
  eprint = {astro-ph/0302208},
  pages = {97},
  issn = {0067-0049},
  doi = {10.1086/377252},
  archiveprefix = {arxiv},
  collaboration = {WMAP}
}

@article{carones2023MultiClusteringNeedletILCCMB,
  author = "Carones, Alessandro and Migliaccio, Marina and Puglisi, Giuseppe and Baccigalupi, Carlo and Marinucci, Domenico and Vittorio, Nicola and Poletti, Davide",
  collaboration = "LiteBIRD",
  title = "{Multi-clustering needlet ILC for CMB B-mode component separation}",
  eprint = "2212.04456",
  archivePrefix = "arXiv",
  primaryClass = "astro-ph.CO",
  doi = "10.1093/mnras/stad2423",
  journal = "Mon. Not. Roy. Astron. Soc.",
  volume = "525",
  number = "2",
  pages = "3117--3135",
  year = "2023"
}

@article{delabrouille2003MultidetectorMulticomponentSpectral,
  title = {Multidetector Multicomponent Spectral Matching and Applications for Cosmic Microwave Background Data Analysis},
  author = {Delabrouille, J. and Cardoso, J.-F. and Patanchon, G.},
  year = {2003},
  journal = {Monthly Notices of the Royal Astronomical Society},
  volume = {346},
  number = {4},
  eprint = {astro-ph/0211504},
  pages = {1089},
  issn = {0035-8711},
  doi = {10.1111/j.1365-2966.2003.07069.x},
  archiveprefix = {arxiv}
}

@article{delabrouille2009FullSkyLow,
  title = {A Full Sky, Low Foreground, High Resolution {{CMB}} Map from {{WMAP}}},
  author = {Delabrouille, J. and Cardoso, J. -F. and Le Jeune, M. and Betoule, M. and Fay, G. and Guilloux, F.},
  year = {2009},
  journal = {Astronomy and Astrophysics},
  volume = {493},
  eprint = {0807.0773},
  primaryclass = {astro-ph},
  pages = {835},
  issn = {0004-6361},
  doi = {10.1051/0004-6361:200810514},
  archiveprefix = {arxiv}
}

@article{rosenberg2022CMBPowerSpectraCosmological,
  title = {{{CMB}} Power Spectra and Cosmological Parameters from {{Planck PR4}} with {{CamSpec}}},
  author = {Rosenberg, Erik and Gratton, Steven and Efstathiou, George},
  year = {2022},
  month = nov,
  journal = {Monthly Notices of the Royal Astronomical Society},
  volume = {517},
  number = {3},
  eprint = {2205.10869},
  primaryclass = {astro-ph},
  pages = {4620--4636},
  issn = {0035-8711, 1365-2966},
  doi = {10.1093/mnras/stac2744},
  archiveprefix = {arxiv}
}

@article{mangilli2008ImpactCosmicNeutrinosGravitationalwave,
  title = {The Impact of Cosmic Neutrinos on the Gravitational-Wave Background},
  author = {Mangilli, A. and Bartolo, N. and Matarrese, S. and Riotto, A.},
  year = {2008},
  journal = {Physical Review D},
  volume = {78},
  number = {8},
  eprint = {0805.3234},
  primaryclass = {astro-ph},
  pages = {083517},
  issn = {1550-7998, 1550-2368},
  doi = {10.1103/PhysRevD.78.083517},
  archiveprefix = {arxiv}
}

@article{Pagano_2020,
   title={Reionization optical depth determination from Planck HFI data with ten percent accuracy},
   volume={635},
   ISSN={1432-0746},
   DOI={10.1051/0004-6361/201936630},
   journal={Astronomy \& Astrophysics},
   publisher={EDP Sciences},
   author={Pagano, L. and Delouis, J.-M. and Mottet, S. and Puget, J.-L. and Vibert, L.},
   year={2020},
   month=mar, pages={A99}
}

@article{Natale_2020,
   title={A novel CMB polarization likelihood package for large angular scales built from combined WMAP and Planck LFI legacy maps},
   volume={644},
   ISSN={1432-0746},
   DOI={10.1051/0004-6361/202038508},
   journal={Astronomy \& Astrophysics},
   publisher={EDP Sciences},
   author={Natale, U. and Pagano, L. and Lattanzi, M. and Migliaccio, M. and Colombo, L. P. and Gruppuso, A. and Natoli, P. and Polenta, G.},
   year={2020},
   month=nov, pages={A32}
}

@article{Lattanzi_2017,
   title={On the impact of large angle CMB polarization data on cosmological parameters},
   volume={2017},
   ISSN={1475-7516},
   DOI={10.1088/1475-7516/2017/02/041},
   number={02},
   journal={Journal of Cosmology and Astroparticle Physics},
   publisher={IOP Publishing},
   author={Lattanzi, Massimiliano and Burigana, Carlo and Gerbino, Martina and Gruppuso, Alessandro and Mandolesi, Nazzareno and Natoli, Paolo and Polenta, Gianluca and Salvati, Laura and Trombetti, Tiziana},
   year={2017},
   month=feb, pages={041–041}
}

@article{de_Belsunce_2021,
   title={Inference of the optical depth to reionization from low multipole temperature and polarization Planck data},
   volume={507},
   ISSN={1365-2966},
   DOI={10.1093/mnras/stab2215},
   number={1},
   journal={Monthly Notices of the Royal Astronomical Society},
   publisher={Oxford University Press (OUP)},
   author={de Belsunce, Roger and Gratton, Steven and Coulton, William and Efstathiou, George},
   year={2021},
   month=aug, pages={1072–1091}
}

@article{Percival_2006,
   title={Likelihood techniques for the combined analysis of CMB temperature and polarization power spectra},
   volume={372},
   ISSN={1365-2966},
   DOI={10.1111/j.1365-2966.2006.10910.x},
   number={3},
   journal={Monthly Notices of the Royal Astronomical Society},
   publisher={Oxford University Press (OUP)},
   author={Percival, W. J. and Brown, M. L.},
   year={2006},
   month=nov, pages={1104–1116}
}

@article{Tegmark_2001,
   title={How to measure CMB polarization power spectra without losing information},
   volume={64},
   ISSN={1089-4918},
   DOI={10.1103/physrevd.64.063001},
   number={6},
   journal={Physical Review D},
   publisher={American Physical Society (APS)},
   author={Tegmark, Max and de Oliveira-Costa, Angelica},
   year={2001},
   month=aug
}

@article{galloniUpdatedConstraintsAmplitude2022,
    author = "Galloni, Giacomo and Bartolo, Nicola and Matarrese, Sabino and Migliaccio, Marina and Ricciardone, Angelo and Vittorio, Nicola",
    title = "{Updated constraints on amplitude and tilt of the tensor primordial spectrum}",
    eprint = "2208.00188",
    archivePrefix = "arXiv",
    primaryClass = "astro-ph.CO",
    doi = "10.1088/1475-7516/2023/04/062",
    journal = "JCAP",
    volume = "04",
    pages = "062",
    year = "2023"
}

@article{galloni2024RobustConstraintsTensorPerturbations,
    author = "Galloni, Giacomo and Henrot-Versill\'e, Sophie and Tristram, Matthieu",
    title = "{Robust constraints on tensor perturbations from cosmological data: A comparative analysis from Bayesian and frequentist perspectives}",
    eprint = "2405.04455",
    archivePrefix = "arXiv",
    primaryClass = "astro-ph.CO",
    doi = "10.1103/PhysRevD.110.063511",
    journal = "Phys. Rev. D",
    volume = "110",
    number = "6",
    pages = "063511",
    year = "2024"
}

@article{Tegmark:1994we,
    author = "Tegmark, Max",
    title = "{A Method for extracting maximum resolution power spectra from microwave sky maps}",
    eprint = "astro-ph/9412064",
    archivePrefix = "arXiv",
    reportNumber = "MPI-PHT-94-90",
    doi = "10.1093/mnras/280.1.299",
    journal = "Mon. Not. Roy. Astron. Soc.",
    volume = "280",
    pages = "299--308",
    year = "1996"
}

@article{Bond:1998zw,
    author = "Bond, J. R. and Jaffe, Andrew H. and Knox, L.",
    title = "{Estimating the power spectrum of the cosmic microwave background}",
    eprint = "astro-ph/9708203",
    archivePrefix = "arXiv",
    doi = "10.1103/PhysRevD.57.2117",
    journal = "Phys. Rev. D",
    volume = "57",
    pages = "2117--2137",
    year = "1998"
}

@article{Efstathiou:2003dj,
    author = "Efstathiou, G.",
    title = "{Myths and truths concerning estimation of power spectra}",
    eprint = "astro-ph/0307515",
    archivePrefix = "arXiv",
    doi = "10.1111/j.1365-2966.2004.07530.x",
    journal = "Mon. Not. Roy. Astron. Soc.",
    volume = "349",
    pages = "603",
    year = "2004"
}

@article{Remazeilles2024MappingHotGasLiteBIRD,
       author = {{Remazeilles}, M. and {Douspis}, M. and {Rubi{\~n}o-Mart{\'\i}n}, J.~A. and {Banday}, A.~J. and {Chluba}, J. and {de Bernardis}, P. and {De Petris}, M. and {Hern{\'a}ndez-Monteagudo}, C. and {Luzzi}, G. and {Macias-Perez}, J. and {Masi}, S. and {Namikawa}, T. and {Salvati}, L. and {Tanimura}, H. and {Aizawa}, K. and {Anand}, A. and {Aumont}, J. and {Baccigalupi}, C. and {Ballardini}, M. and {Barreiro}, R.~B. and {Bartolo}, N. and {Basak}, S. and {Bersanelli}, M. and {Blinov}, D. and {Bortolami}, M. and {Brinckmann}, T. and {Calabrese}, E. and {Campeti}, P. and {Carinos}, E. and {Carones}, A. and {Casas}, F.~J. and {Cheung}, K. and {Clermont}, L. and {Columbro}, F. and {Coppolecchia}, A. and {Cuttaia}, F. and {de Haan}, T. and {de la Hoz}, E. and {Della Torre}, S. and {Diego-Palazuelos}, P. and {D'Alessandro}, G. and {Eriksen}, H.~K. and {Finelli}, F. and {Fuskeland}, U. and {Galloni}, G. and {Galloway}, M. and {Gervasi}, M. and {G{\'e}nova-Santos}, R.~T. and {Ghigna}, T. and {Giardiello}, S. and {Gimeno-Amo}, C. and {Gjerl{\o}w}, E. and {Gonz{\'a}lez Gonz{\'a}lez}, R. and {Gruppuso}, A. and {Hazumi}, M. and {Henrot-Versill{\'e}}, S. and {Hergt}, L.~T. and {Herranz}, D. and {Kohri}, K. and {Komatsu}, E. and {Lamagna}, L. and {Lattanzi}, M. and {Leloup}, C. and {Levrier}, F. and {Lonappan}, A.~I. and {L{\'o}pez-Caniego}, M. and {Maffei}, B. and {Mart{\'\i}nez-Gonz{\'a}lez}, E. and {Matarrese}, S. and {Matsumura}, T. and {Micheli}, S. and {Migliaccio}, M. and {Monelli}, M. and {Montier}, L. and {Morgante}, G. and {Nagano}, Y. and {Nagata}, R. and {Novelli}, A. and {Omae}, R. and {Pagano}, L. and {Paoletti}, D. and {Pavlidou}, V. and {Piacentini}, F. and {Pinchera}, M. and {Polenta}, G. and {Porcelli}, L. and {Ritacco}, A. and {Ruiz-Granda}, M. and {Sakurai}, Y. and {Scott}, D. and {Shiraishi}, M. and {Stever}, S.~L. and {Sullivan}, R.~M. and {Takase}, Y. and {Tassis}, K. and {Terenzi}, L. and {Tomasi}, M. and {Tristram}, M. and {Vacher}, L. and {van Tent}, B. and {Vielva}, P. and {Wehus}, I.~K. and {Westbrook}, B. and {Weymann-Despres}, G. and {Wollack}, E.~J. and {Zannoni}, M. and {Zhou}, Y. and {LiteBIRD Collaboration}},
        title = "{LiteBIRD science goals and forecasts. Mapping the hot gas in the Universe}",
      journal = {JCAP},
         year = 2024,
        month = dec,
       volume = {2024},
       number = {12},
          eid = {026},
        pages = {026},
          doi = {10.1088/1475-7516/2024/12/026},
archivePrefix = {arXiv},
       eprint = {2407.17555},
 primaryClass = {astro-ph.CO},
}

@article{Fuskeland2023ExtentedFrequencyConfigurationLiteBIRD,
       author = {{Fuskeland}, U. and {Aumont}, J. and {Aurlien}, R. and {Baccigalupi}, C. and {Banday}, A.~J. and {Eriksen}, H.~K. and {Errard}, J. and {G{\'e}nova-Santos}, R.~T. and {Hasebe}, T. and {Hubmayr}, J. and {Imada}, H. and {Krachmalnicoff}, N. and {Lamagna}, L. and {Pisano}, G. and {Poletti}, D. and {Remazeilles}, M. and {Thompson}, K.~L. and {Vacher}, L. and {Wehus}, I.~K. and {Azzoni}, S. and {Ballardini}, M. and {Barreiro}, R.~B. and {Bartolo}, N. and {Basyrov}, A. and {Beck}, D. and {Bersanelli}, M. and {Bortolami}, M. and {Brilenkov}, M. and {Calabrese}, E. and {Carones}, A. and {Casas}, F.~J. and {Cheung}, K. and {Chluba}, J. and {Clark}, S.~E. and {Clermont}, L. and {Columbro}, F. and {Coppolecchia}, A. and {D'Alessandro}, G. and {de Bernardis}, P. and {de Haan}, T. and {de la Hoz}, E. and {De Petris}, M. and {Della Torre}, S. and {Diego-Palazuelos}, P. and {Finelli}, F. and {Franceschet}, C. and {Galloni}, G. and {Galloway}, M. and {Gerbino}, M. and {Gervasi}, M. and {Ghigna}, T. and {Giardiello}, S. and {Gjerl{\o}w}, E. and {Gruppuso}, A. and {Hargrave}, P. and {Hattori}, M. and {Hazumi}, M. and {Hergt}, L.~T. and {Herman}, D. and {Herranz}, D. and {Hivon}, E. and {Hoang}, T.~D. and {Kohri}, K. and {Lattanzi}, M. and {Lee}, A.~T. and {Leloup}, C. and {Levrier}, F. and {Lonappan}, A.~I. and {Luzzi}, G. and {Maffei}, B. and {Mart{\'\i}nez-Gonz{\'a}lez}, E. and {Masi}, S. and {Matarrese}, S. and {Matsumura}, T. and {Migliaccio}, M. and {Montier}, L. and {Morgante}, G. and {Mot}, B. and {Mousset}, L. and {Nagata}, R. and {Namikawa}, T. and {Nati}, F. and {Natoli}, P. and {Nerval}, S. and {Novelli}, A. and {Pagano}, L. and {Paiella}, A. and {Paoletti}, D. and {Pascual-Cisneros}, G. and {Patanchon}, G. and {Pelgrims}, V. and {Piacentini}, F. and {Piccirilli}, G. and {Polenta}, G. and {Puglisi}, G. and {Raffuzzi}, N. and {Ritacco}, A. and {Rubino-Martin}, J.~A. and {Savini}, G. and {Scott}, D. and {Sekimoto}, Y. and {Shiraishi}, M. and {Signorelli}, G. and {Stever}, S.~L. and {Stutzer}, N. and {Sullivan}, R.~M. and {Takakura}, H. and {Terenzi}, L. and {Thommesen}, H. and {Tristram}, M. and {Tsuji}, M. and {Vielva}, P. and {Weller}, J. and {Westbrook}, B. and {Weymann-Despres}, G. and {Wollack}, E.~J. and {Zannoni}, M.},
        title = "{Tensor-to-scalar ratio forecasts for extended LiteBIRD frequency configurations}",
      journal = {\aap},
         year = 2023,
        month = aug,
       volume = {676},
          eid = {A42},
        pages = {A42},
          doi = {10.1051/0004-6361/202346155},
archivePrefix = {arXiv},
       eprint = {2302.05228},
 primaryClass = {astro-ph.CO},
}

@article{Knox:1995dq,
    author = "Knox, LLoyd",
    title = "{Determination of inflationary observables by cosmic microwave background anisotropy experiments}",
    eprint = "astro-ph/9504054",
    archivePrefix = "arXiv",
    reportNumber = "FERMILAB-PUB-95-008-A",
    doi = "10.1103/PhysRevD.52.4307",
    journal = "Phys. Rev. D",
    volume = "52",
    pages = "4307--4318",
    year = "1995"
}

@article{bicepPlanck2015jointAnalysis,
       author = {{BICEP2/Keck Collaboration} and {Planck Collaboration} and {Ade}, P.~A.~R. and {Aghanim}, N. and {Ahmed}, Z. and {Aikin}, R.~W. and {Alexander}, K.~D. and {Arnaud}, M. and {Aumont}, J. and {Baccigalupi}, C. and {Banday}, A.~J. and {Barkats}, D. and {Barreiro}, R.~B. and {Bartlett}, J.~G. and {Bartolo}, N. and {Battaner}, E. and {Benabed}, K. and {Beno{\^\i}t}, A. and {Benoit-L{\'e}vy}, A. and {Benton}, S.~J. and {Bernard}, J. -P. and {Bersanelli}, M. and {Bielewicz}, P. and {Bischoff}, C.~A. and {Bock}, J.~J. and {Bonaldi}, A. and {Bonavera}, L. and {Bond}, J.~R. and {Borrill}, J. and {Bouchet}, F.~R. and {Boulanger}, F. and {Brevik}, J.~A. and {Bucher}, M. and {Buder}, I. and {Bullock}, E. and {Burigana}, C. and {Butler}, R.~C. and {Buza}, V. and {Calabrese}, E. and {Cardoso}, J. -F. and {Catalano}, A. and {Challinor}, A. and {Chary}, R. -R. and {Chiang}, H.~C. and {Christensen}, P.~R. and {Colombo}, L.~P.~L. and {Combet}, C. and {Connors}, J. and {Couchot}, F. and {Coulais}, A. and {Crill}, B.~P. and {Curto}, A. and {Cuttaia}, F. and {Danese}, L. and {Davies}, R.~D. and {Davis}, R.~J. and {de Bernardis}, P. and {De Rosa}, A. and {de Zotti}, G. and {Delabrouille}, J. and {Delouis}, J. -M. and {D{\'e}sert}, F. -X. and {Dickinson}, C. and {Diego}, J.~M. and {Dole}, H. and {Donzelli}, S. and {Dor{\'e}}, O. and {Douspis}, M. and {Dowell}, C.~D. and {Duband}, L. and {Ducout}, A. and {Dunkley}, J. and {Dupac}, X. and {Dvorkin}, C. and {Efstathiou}, G. and {Elsner}, F. and {En{\ss}lin}, T.~A. and {Eriksen}, H.~K. and {Falgarone}, E. and {Filippini}, J.~P. and {Finelli}, F. and {Fliescher}, S. and {Forni}, O. and {Frailis}, M. and {Fraisse}, A.~A. and {Franceschi}, E. and {Frejsel}, A. and {Galeotta}, S. and {Galli}, S. and {Ganga}, K. and {Ghosh}, T. and {Giard}, M. and {Gjerl{\o}w}, E. and {Golwala}, S.~R. and {Gonz{\'a}lez-Nuevo}, J. and {G{\'o}rski}, K.~M. and {Gratton}, S. and {Gregorio}, A. and {Gruppuso}, A. and {Gudmundsson}, J.~E. and {Halpern}, M. and {Hansen}, F.~K. and {Hanson}, D. and {Harrison}, D.~L. and {Hasselfield}, M. and {Helou}, G. and {Henrot-Versill{\'e}}, S. and {Herranz}, D. and {Hildebrandt}, S.~R. and {Hilton}, G.~C. and {Hivon}, E. and {Hobson}, M. and {Holmes}, W.~A. and {Hovest}, W. and {Hristov}, V.~V. and {Huffenberger}, K.~M. and {Hui}, H. and {Hurier}, G. and {Irwin}, K.~D. and {Jaffe}, A.~H. and {Jaffe}, T.~R. and {Jewell}, J. and {Jones}, W.~C. and {Juvela}, M. and {Karakci}, A. and {Karkare}, K.~S. and {Kaufman}, J.~P. and {Keating}, B.~G. and {Kefeli}, S. and {Keih{\"a}nen}, E. and {Kernasovskiy}, S.~A. and {Keskitalo}, R. and {Kisner}, T.~S. and {Kneissl}, R. and {Knoche}, J. and {Knox}, L. and {Kovac}, J.~M. and {Krachmalnicoff}, N. and {Kunz}, M. and {Kuo}, C.~L. and {Kurki-Suonio}, H. and {Lagache}, G. and {L{\"a}hteenm{\"a}ki}, A. and {Lamarre}, J. -M. and {Lasenby}, A. and {Lattanzi}, M. and {Lawrence}, C.~R. and {Leitch}, E.~M. and {Leonardi}, R. and {Levrier}, F. and {Lewis}, A. and {Liguori}, M. and {Lilje}, P.~B. and {Linden-V{\o}rnle}, M. and {L{\'o}pez-Caniego}, M. and {Lubin}, P.~M. and {Lueker}, M. and {Mac{\'\i}as-P{\'e}rez}, J.~F. and {Maffei}, B. and {Maino}, D. and {Mandolesi}, N. and {Mangilli}, A. and {Maris}, M. and {Martin}, P.~G. and {Mart{\'\i}nez-Gonz{\'a}lez}, E. and {Masi}, S. and {Mason}, P. and {Matarrese}, S. and {Megerian}, K.~G. and {Meinhold}, P.~R. and {Melchiorri}, A. and {Mendes}, L. and {Mennella}, A. and {Migliaccio}, M. and {Mitra}, S. and {Miville-Desch{\^e}nes}, M. -A. and {Moneti}, A. and {Montier}, L. and {Morgante}, G. and {Mortlock}, D. and {Moss}, A. and {Munshi}, D. and {Murphy}, J.~A. and {Naselsky}, P. and {Nati}, F. and {Natoli}, P. and {Netterfield}, C.~B. and {Nguyen}, H.~T. and {N{\o}rgaard-Nielsen}, H.~U. and {Noviello}, F. and {Novikov}, D. and {Novikov}, I. and {O'Brient}, R. and {Ogburn}, R.~W. and {Orlando}, A. and {Pagano}, L. and {Pajot}, F. and {Paladini}, R. and {Paoletti}, D. and {Partridge}, B.},
        title = "{Joint Analysis of BICEP2/Keck Array and Planck Data}",
      journal = {Physical Review Letters},
         year = 2015,
        month = mar,
       volume = {114},
       number = {10},
          eid = {101301},
        pages = {101301},
          doi = {10.1103/PhysRevLett.114.101301},
archivePrefix = {arXiv},
       eprint = {1502.00612},
 primaryClass = {astro-ph.CO},
}

@phdthesis{Vanneste:2019jnz,
    author = "Vanneste, Sylvain",
    title = "{Constraints on primordial gravitational waves from the large scales CMB data}",
    reportNumber = "tel-02426412, 2019SACLS314",
    school = "U. Paris-Saclay",
    year = "2019",
    url = "https://theses.fr/2019SACLS314",
}

@article{Delouis2019Sroll2,
    author = "Delouis, J. -M. and Pagano, L. and Mottet, S. and Puget, J. -L. and Vibert, L.",
    title = "{SRoll2: an improved mapmaking approach to reduce large-scale systematic effects in the Planck High Frequency Instrument legacy maps}",
    eprint = "1901.11386",
    archivePrefix = "arXiv",
    primaryClass = "astro-ph.CO",
    doi = "10.1051/0004-6361/201834882",
    journal = "Astron. Astrophys.",
    volume = "629",
    pages = "A38",
    year = "2019"
}

@article{planck2016reductionLargeScaleSystematics,
    author={Aghanim, N. and Ashdown, M. and Aumont, J. and Baccigalupi, C. and Ballardini, M. and Banday, A. J. and Barreiro, R. B. and Bartolo, N. and Basak, S. and Battye, R. and Benabed, K. and Bernard, J.-P. and Bersanelli, M. and Bielewicz, P. and Bock, J. J. and Bonaldi, A. and Bonavera, L. and Bond, J. R. and Borrill, J. and Bouchet, F. R. and Boulanger, F. and Bucher, M. and Burigana, C. and Butler, R. C. and Calabrese, E. and Cardoso, J.-F. and Carron, J. and Challinor, A. and Chiang, H. C. and Colombo, L. P. L. and Combet, C. and Comis, B. and Coulais, A. and Crill, B. P. and Curto, A. and Cuttaia, F. and Davis, R. J. and de Bernardis, P. and de Rosa, A. and de Zotti, G. and Delabrouille, J. and Delouis, J.-M. and Di Valentino, E. and Dickinson, C. and Diego, J. M. and Doré, O. and Douspis, M. and Ducout, A. and Dupac, X. and Efstathiou, G. and Elsner, F. and Enßlin, T. A. and Eriksen, H. K. and Falgarone, E. and Fantaye, Y. and Finelli, F. and Forastieri, F. and Frailis, M. and Fraisse, A. A. and Franceschi, E. and Frolov, A. and Galeotta, S. and Galli, S. and Ganga, K. and Génova-Santos, R. T. and Gerbino, M. and Ghosh, T. and González-Nuevo, J. and Górski, K. M. and Gratton, S. and Gruppuso, A. and Gudmundsson, J. E. and Hansen, F. K. and Helou, G. and Henrot-Versillé, S. and Herranz, D. and Hivon, E. and Huang, Z. and Ilić, S. and Jaffe, A. H. and Jones, W. C. and Keihänen, E. and Keskitalo, R. and Kisner, T. S. and Knox, L. and Krachmalnicoff, N. and Kunz, M. and Kurki-Suonio, H. and Lagache, G. and Lamarre, J.-M. and Langer, M. and Lasenby, A. and Lattanzi, M. and Lawrence, C. R. and Le Jeune, M. and Leahy, J. P. and Levrier, F. and Liguori, M. and Lilje, P. B. and López-Caniego, M. and Ma, Y.-Z. and Macías-Pérez, J. F. and Maggio, G. and Mangilli, A. and Maris, M. and Martin, P. G. and Martínez-González, E. and Matarrese, S. and Mauri, N. and McEwen, J. D. and Meinhold, P. R. and Melchiorri, A. and Mennella, A. and Migliaccio, M. and Miville-Deschênes, M.-A. and Molinari, D. and Moneti, A. and Montier, L. and Morgante, G. and Moss, A. and Mottet, S. and Naselsky, P. and Natoli, P. and Oxborrow, C. A. and Pagano, L. and Paoletti, D. and Partridge, B. and Patanchon, G. and Patrizii, L. and Perdereau, O. and Perotto, L. and Pettorino, V. and Piacentini, F. and Plaszczynski, S. and Polastri, L. and Polenta, G. and Puget, J.-L. and Rachen, J. P. and Racine, B. and Reinecke, M. and Remazeilles, M. and Renzi, A. and Rocha, G. and Rossetti, M. and Roudier, G. and Rubiño-Martín, J. A. and Ruiz-Granados, B. and Salvati, L. and Sandri, M. and Savelainen, M. and Scott, D. and Sirri, G. and Sunyaev, R. and Suur-Uski, A.-S. and Tauber, J. A. and Tenti, M. and Toffolatti, L. and Tomasi, M. and Tristram, M. and Trombetti, T. and Valiviita, J. and Van Tent, F. and Vibert, L. and Vielva, P. and Villa, F. and Vittorio, N. and Wandelt, B. D. and Watson, R. and Wehus, I. K. and White, M. and Zacchei, A. and Zonca, A.},
    collaboration = "Planck",
    title = "{Planck intermediate results. XLVI. Reduction of large-scale systematic effects in HFI polarization maps and estimation of the reionization optical depth}",
    eprint = "1605.02985",
    archivePrefix = "arXiv",
    primaryClass = "astro-ph.CO",
    doi = "10.1051/0004-6361/201628890",
    journal = "Astron. Astrophys.",
    volume = "596",
    pages = "A107",
    year = "2016"
}

@ARTICLE{benabed2009teasing,
       author = {{Benabed}, K. and {Cardoso}, J. -F. and {Prunet}, S. and {Hivon}, E.},
        title = "{TEASING: a fast and accurate approximation for the low multipole likelihood of the cosmic microwave background temperature}",
      journal = {\mnras},
         year = 2009,
        month = nov,
       volume = {400},
       number = {1},
        pages = {219-227},
          doi = {10.1111/j.1365-2966.2009.15202.x},
archivePrefix = {arXiv},
       eprint = {0901.4537},
 primaryClass = {astro-ph.CO},
       adsurl = {https://ui.adsabs.harvard.edu/abs/2009MNRAS.400..219B}
}

@article{Jewell:2008hg,
    author = "Jewell, J. B. and Eriksen, H. K. and Wandelt, B. D. and O'Dwyer, I. J. and Huey, G. and Gorski, K. M.",
    title = "{A Markov Chain Monte Carlo Algorithm for analysis of low signal-to-noise CMB data}",
    eprint = "0807.0624",
    archivePrefix = "arXiv",
    primaryClass = "astro-ph",
    doi = "10.1088/0004-637X/697/1/258",
    journal = "Astrophys. J.",
    volume = "697",
    pages = "258--268",
    year = "2009"
}

@article{Taylor:2007sj,
    author = "Taylor, J. F. and Ashdown, M. A. J. and Hobson, M. P.",
    title = "{Fast optimal CMB power spectrum estimation with Hamiltonian sampling}",
    eprint = "0708.2989",
    archivePrefix = "arXiv",
    primaryClass = "astro-ph",
    doi = "10.1111/j.1365-2966.2008.13630.x",
    journal = "Mon. Not. Roy. Astron. Soc.",
    volume = "389",
    pages = "1284",
    year = "2008"
}

@article{Eriksen:2004ss,
    author = "Eriksen, Hans Kristian and O'Dwyer, I. J. and Jewell, J. B. and Wandelt, B. D. and Larson, D. L. and Gorski, K. M. and Levin, S. and Banday, A. J. and Lilje, P. B.",
    title = "{Power spectrum estimation from high-resolution maps by Gibbs sampling}",
    eprint = "astro-ph/0407028",
    archivePrefix = "arXiv",
    doi = "10.1086/425219",
    journal = "Astrophys. J. Suppl.",
    volume = "155",
    pages = "227--241",
    year = "2004"
}

@article{deBelsunce:2022yll,
    author = "de Belsunce, Roger and Gratton, Steven and Efstathiou, George",
    title = "{B-mode constraints from Planck low-multipole polarization data}",
    eprint = "2207.04903",
    archivePrefix = "arXiv",
    primaryClass = "astro-ph.CO",
    doi = "10.1093/mnras/stac3321",
    journal = "Mon. Not. Roy. Astron. Soc.",
    volume = "518",
    number = "3",
    pages = "3675--3684",
    year = "2022"
}

@article{Gorski:1994uu,
    author = "Gorski, K. M. and Hinshaw, G. and Banday, a. J. and Bennett, C. L. and Wright, E. L. and Kogut, a. and Smoot, George F. Smoot and Lubin, P.",
    title = "{On determining the spectrum of primordial inhomogeneity from the COBE dmr sky maps. 2. Results of two year data analysis}",
    eprint = "astro-ph/9403067",
    archivePrefix = "arXiv",
    doi = "10.1086/187445",
    journal = "Astrophys. J. Lett.",
    volume = "430",
    pages = "L89--L92",
    year = "1994"
}

@article{Gorski:1994ye,
    author = "Gorski, Krzysztof M.",
    title = "{On Determining the spectrum of primordial inhomogeneity from the Cobe DMR sky maps. 1. Method}",
    eprint = "astro-ph/9403066",
    archivePrefix = "arXiv",
    reportNumber = "COBE-94-07",
    doi = "10.1086/187444",
    journal = "Astrophys. J. Lett.",
    volume = "430",
    pages = "L85",
    year = "1994"
}

@article{Gorski:1996cf,
    author = "Gorski, K. M. and Banday, A. J. and Bennett, C. L. and Hinshaw, G. and Kogut, A. and Smoot, George F. and Wright, E. L.",
    title = "{Power spectrum of primordial inhomogeneity determined from the four year COBE DMR sky maps}",
    eprint = "astro-ph/9601063",
    archivePrefix = "arXiv",
    reportNumber = "COBE-PREPRINT-96-03",
    doi = "10.1086/310077",
    journal = "Astrophys. J. Lett.",
    volume = "464",
    pages = "L11",
    year = "1996"
}

@inproceedings{Gorski:1996ti,
    author = "G{\'o}rski, Krzysztof M.",
    title = "{Cosmic microwave background anisotropy in the COBE DMR 4-year sky maps}",
    booktitle = "{31st Rencontres de Moriond: Microwave Background Anisotropies}",
    eprint = "astro-ph/9701191",
    archivePrefix = "arXiv",
    pages = "77--84",
    year = "1996"
}

@article{smoot1992cosmicBackgroundExplorer,
       author = {{Smoot}, G.~F. and {Bennett}, C.~L. and {Kogut}, A. and {Wright}, E.~L. and {Aymon}, J. and {Boggess}, N.~W. and {Cheng}, E.~S. and {de Amici}, G. and {Gulkis}, S. and {Hauser}, M.~G. and {Hinshaw}, G. and {Jackson}, P.~D. and {Janssen}, M. and {Kaita}, E. and {Kelsall}, T. and {Keegstra}, P. and {Lineweaver}, C. and {Loewenstein}, K. and {Lubin}, P. and {Mather}, J. and {Meyer}, S.~S. and {Moseley}, S.~H. and {Murdock}, T. and {Rokke}, L. and {Silverberg}, R.~F. and {Tenorio}, L. and {Weiss}, R. and {Wilkinson}, D.~T.},
        title = "{Structure in the COBE Differential Microwave Radiometer First-Year Maps}",
      journal = {\apjl},
         year = 1992,
        month = sep,
       volume = {396},
        pages = {L1},
          doi = {10.1086/186504},
}

\end{document}